\documentclass[aps,pra,preprint,bibnotes,superscriptaddress,amsmath,amssymb,longbibliography,floatfix]{revtex4-1}
\usepackage[T1]{fontenc}
\usepackage{amsbsy,bm,bbold}
\usepackage{graphicx,color,xcolor,epsfig,rotate}
\usepackage{gensymb}
\usepackage{blindtext}
\usepackage{dcolumn}
\usepackage[normalem]{ulem}
\usepackage[left]{lineno}
\definecolor{magenta}{cmyk}{ 0, 1, 0,0}

\begin{document}

\title{Tomonaga-Luttinger Liquid Behavior and Spinon Confinement in YbAlO$_3$ }

\author{L.~S.~Wu}
\thanks{These authors contributed equally to this work}
\affiliation{Neutron Scattering Division, Oak Ridge National Laboratory, Oak Ridge, TN 37831, USA}
\affiliation{Department of Physics, Southern University of Science and Technology, Shenzhen 518055, China}
\email[Corresponding author: ]{wuls@sustc.edu.cn}

\author{S.~E.~Nikitin}
\thanks{These authors contributed equally to this work}
\affiliation{Max Planck Institute for Chemical Physics of Solids, N\"{o}thnitzer Str. 40, D-01187 Dresden, Germany}
\affiliation{Institut f{\"u}r Festk{\"o}rper- und Materialphysik, Technische Universit{\"a}t Dresden, D-01069 Dresden, Germany}

\author{Z.~Wang}
\thanks{These authors contributed equally to this work}
\affiliation{Department of Physics and Astronomy, The University of Tennessee, Knoxville, TN 37996, USA}

\author{W.~Zhu}
\affiliation{Westlake Institute of Advanced Study, Hangzhou, 310024, P.~R.~China}
\affiliation{Theoretical Division, T-4 and CNLS, Los Alamos National Laboratory, Los Alamos, NM 87545, USA}

\author{C.~D.~Batista}
\affiliation{Department of Physics and Astronomy, The University of Tennessee, Knoxville, TN 37996, USA}
\affiliation{Shull-Wollan Center, Oak Ridge National Laboratory, Oak Ridge, TN 37831, USA}

\author{A.~M.~Tsvelik}
\affiliation{Condensed Matter Physics and Materials Science Division, Brookhaven National Laboratory, Upton, NY 11973, USA}

\author{A.~M.~Samarakoon}
\affiliation{Neutron Scattering Division, Oak Ridge National Laboratory, Oak Ridge, TN 37831, USA}

\author{D.~A.~Tennant}
\affiliation{Materials Science and Technology Division, Oak Ridge National Laboratory, Oak Ridge, TN 37831, U.S.A.}
\affiliation{Shull-Wollan Center, Oak Ridge National Laboratory, Oak Ridge, TN 37831, USA}

\author{M.~Brando}
\affiliation{Max Planck Institute for Chemical Physics of Solids, N\"{o}thnitzer Str. 40, D-01187 Dresden, Germany}

\author{L.~Vasylechko}
\affiliation{Lviv Polytechnic National University, 79013 Lviv, Ukraine}

\author{M.~Frontzek}
\affiliation{Neutron Scattering Division, Oak Ridge National Laboratory, Oak Ridge, TN 37831, USA}

\author{A.~T.~Savici}
\affiliation{Neutron Scattering Division, Oak Ridge National Laboratory, Oak Ridge, TN 37831, USA}

\author{G.~Sala}
\affiliation{Neutron Scattering Division, Oak Ridge National Laboratory, Oak Ridge, TN 37831, USA}

\author{G.~Ehlers}
\affiliation{Neutron Technologies Division, Oak Ridge National Laboratory, Oak Ridge, TN 37831, USA}

\author{A.~D.~Christianson}
\affiliation{Materials Science and Technology Division, Oak Ridge National Laboratory, Oak Ridge, TN 37831, U.S.A.}
\affiliation{Neutron Scattering Division, Oak Ridge National Laboratory, Oak Ridge, TN 37831, USA}

\author{M.~D.~Lumsden}
\affiliation{Neutron Scattering Division, Oak Ridge National Laboratory, Oak Ridge, TN 37831, USA}

\author{A.~Podlesnyak}
\email[Corresponding author: ]{podlesnyakaa@ornl.gov}
\affiliation{Neutron Scattering Division, Oak Ridge National Laboratory, Oak Ridge, TN 37831, USA}

\maketitle

{ \bf{ABSTRACT\\ Low dimensional quantum magnets are interesting because of the emerging collective behavior arising from strong quantum fluctuations. The one-dimensional (1D) {\emph S}~=~1/2 Heisenberg antiferromagnet is  a paradigmatic example, whose
low-energy excitations, known as spinons, carry fractional spin {\emph S}~=~1/2. These fractional modes can be  reconfined
by  the application of a staggered magnetic field. Even though considerable progress has been made in the theoretical understanding of such magnets, experimental realizations of this low-dimensional physics are relatively rare. This is particularly true for rare-earth based magnets because of the large effective spin anisotropy  induced by the combination of strong spin-orbit coupling and crystal field splitting. Here, we demonstrate that the rare-earth perovskite YbAlO$_{\mathbf{3}}$ provides a realization of a quantum spin {\emph S}~=~1/2 chain material exhibiting both quantum critical Tomonaga-Luttinger liquid behavior and spinon confinement-deconfinement transitions in different regions of magnetic field-temperature  phase diagram.}}\\

\section*{Introduction}

The spin $S=1/2$ antiferromagnetic Heisenberg Hamiltonian is one of the simplest models of condensed matter physics containing fractional quantum number excitations. The phase diagram of this model includes both quantum critical (for easy-plane anisotropy) and gapped phases (for the easy-axis anisotropy). Spinons are topological excitations which can be pictured as domain walls between different N\'eel ground states of the system (this picture is especially appropriate for the moments with easy-axis anisotropy). In view of its exotic physics, it is  exciting to find realizations of this model and it is particularly exciting to find them in unexpected places. The three-dimensional metallic system Yb$_2$Pt$_2$Pb is a recent example, where the neutron scattering revealed the existence of one-dimensional spinon continuum \cite{Wu2016,Classen2018}.
The unusual properties of Yb$_2$Pt$_2$Pb were initially attributed to its peculiar crystal structure~\cite{Kim2013,Ochiai2011}, but more recent work \cite{Wu2016}  demonstrated  that the peculiar form of exchange interactions in  rare earth ions plays a dominant role. However, the metallic nature of this material complicates the analysis of its magnetic properties.

Here we discuss the insulating analog of Yb$_2$Pt$_2$Pb, namely Yb-based quasi-1D quantum magnet, YbAlO$_3$~\cite{Radhakrishna1981}.
RKKY interactions are not present because YbAlO$_3$ is an insulator. Despite the strong
uniaxial single ion anisotropy of the Yb magnetic moments, our neutron scattering data demonstrates that, unlike  Yb$_2$Pt$_2$Pb, YbAlO$_3$ is described by a nearly isotropic (Heisenberg) intrachain interaction, which results in Tomonaga-Luttinger liquid behavior~\cite{Tomonaga1950,Luttinger1963,Mattis1965} over a finite window of applied magnetic field values. Below the N\'eel temperature $T_{\rm N}=0.88$ K, the ordered moments from neighboring chains produce a staggered molecular field that confines the fractional spinon excitations. An additional advantage of YbAlO$_3$ is that the moments can be saturated with a relatively low magnetic field of 1.1~T, enabling exploration of the entire magnetic field-temperature phase diagram.

\section*{Results}
\subsection*{Magnetization, Single ion anisotropy and Two magnetic sublattices}
YbAlO$_3$ crystallizes in an orthorhombically distorted perovskite structure~\cite{Buryy2010},
with room temperature lattice constants $a=5.126$~{\AA}, $b=5.331$~{\AA}, and $c=7.313$~{\AA} (in conventional $Pbnm$ notation).
The local point-group symmetry of the Yb$^{3+}$ ions splits the eight-fold degenerate $J=7/2$ ($L=3$, $S=1/2$) ground state multiplet ($2J+1=8$) into four doublet states.
The ground doublet state $m_J = \pm 7/2$ with small admixture of other states is well separated from the first excited levels~\cite{Radhakrishna1981}.
This ensures that the low temperature and low field magnetic properties can be described by a pseudo-spin 1/2 model.

\begin{figure}
\centering
\includegraphics[width=0.8\columnwidth]{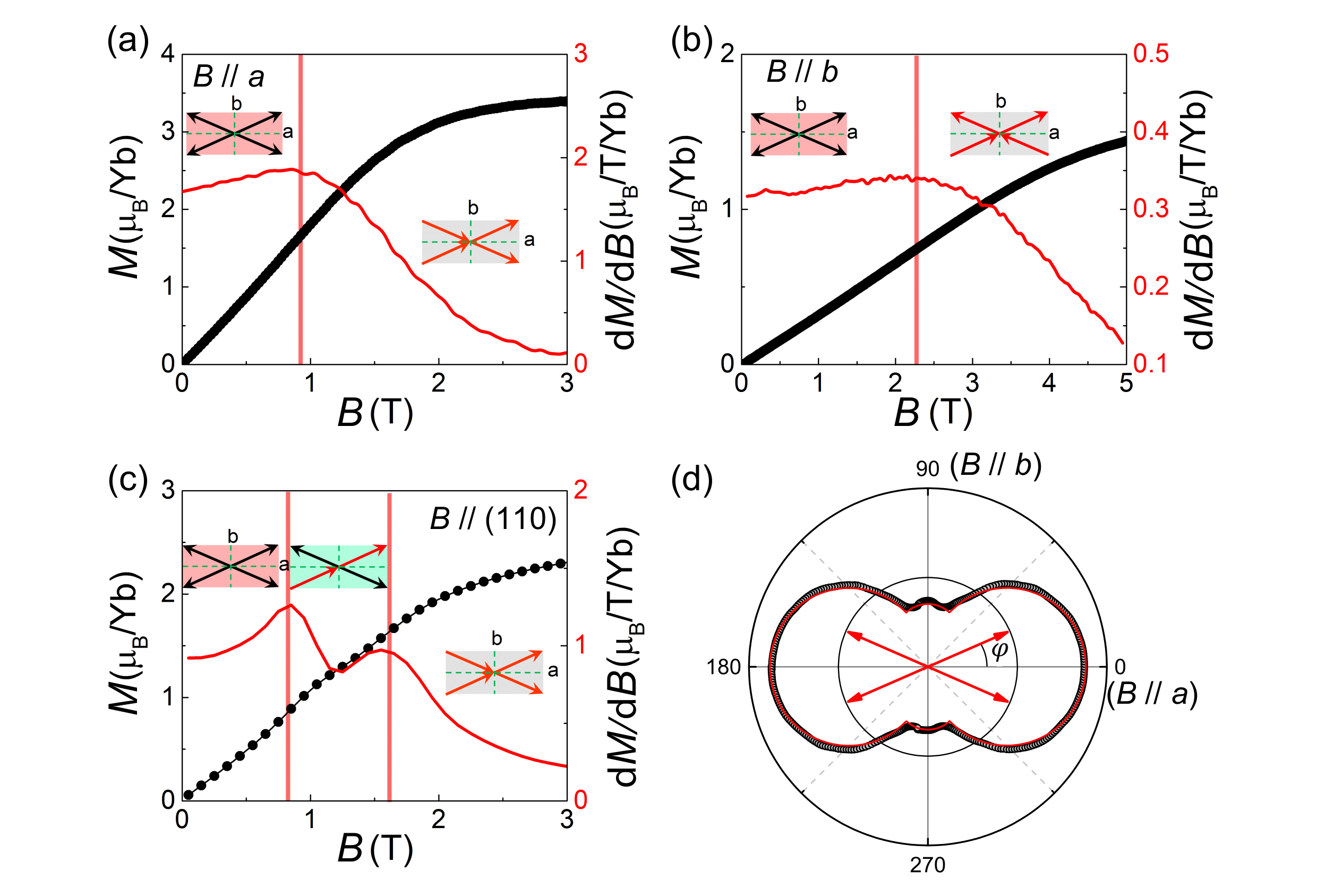}
  \caption{Single ion anisotropy of YbAlO$_3$. (a)-(c) Field dependent magnetization $M$ (black circles) and magnetic susceptibility ${\rm d}M/{\rm d}B$ (red line) of YbAlO$_3$, measured at $T=2$~K with field along the axes $B\parallel a$ , $B\parallel b $ and $B\parallel $~[110], as indicated. The insets are sketches of the magnetic moment configurations in different field ranges. (d) Angle dependent magnetization measured at $T=2$~K and $B=5$~T. Red arrows schematically show a moment configuration of Yb$^{3+}$ at zero field with angle $\varphi=23.5^{\circ}$ between the $a$-axis and the Yb magnetic moment. The solid line is the calculation, as described in the text.}
  \label{MB}
\end{figure}

Crystalline electrical fields (CEF) calculations and magnetization measurements reveal strong uniaxial (Ising-like) single-ion anisotropy, with easy-axis in the $ab$-plane.
Figure~\ref{MB}a-c shows the field dependence of magnetization, $M$, measured at $T=2$~K.
At $B=5$~T, the angle dependent magnetization ({Fig.~\ref{MB}d}) is well described by the function
\begin{equation}
\label{MvsB}
M \approx\dfrac{M_{\rm{s}}}{2}\left(\lvert\cos(\theta-\varphi)\rvert+\lvert\cos(\varphi+\theta)\rvert\right) {\;},
\end{equation}
where $\varphi=23.5^{\circ}$ is the tilting angle of the local easy-axis relative to the $a$-axis, $\theta$ indicates the direction of the applied field in the $ab$-plane, and $M_{\rm{s}}=3.8$~$\mu_{\rm{B}}{\rm{/Yb}}$ is the saturation moment~\cite{Radhakrishna1981}. The effective $g$-factors from magnetization measurements are: $g^{xx} {\simeq}  g^{yy}=0.46$, and $g^{zz}=7.6$~\cite{Radhakrishna1981}, {which are consistent with early studies by M{\"o}ssbauer effect and electron paramagnetic resonance (EPR) measurements~\cite{Bonville1978, Bonville1980}.}

In addition to the $g$-tensor anisotropy, the field dependent magnetization reveals that the coupling between Ising moments with different tilting angles $\pm\varphi$ is weak:
{only one transition is observed for $B\parallel{a}$, $B\parallel{b}$ (Fig.~\ref{MB}a,b), while two successive transitions are found for $B\parallel$~[110] (Fig.~\ref{MB}c), suggesting that magnetic moments with different easy-axis orientation can be individually tuned.}

\subsection*{Model Hamiltonian and zero-field inelastic neutron scattering spectrum}
The most important aspect of the Yb-physics is the exchange interaction. Contrary to naive expectations, the Yb Kramers doublets do not behave as classical Ising spins when they interact with each other. The exchange processes between the pseudospins on the same chain include spin flip terms of quantum origin. The reason for this was described in Ref.~\onlinecite{Wu2016}: the superexchange interaction between total angular momenta of rare earth ions includes matrix elements between all eigenstates, not just between those differing by $\delta J^z =\pm 1$. Being projected on the lowest Kramers doublet, this interaction acquires the familiar $S^aS^a$-form. We remind the reader that spin $S=1/2$ operators {$S^a$} act on the Kramers doublet and, {to a very good approximation, only $S^z$ is linearly related to the total angular moment: $J^z = g^{zz} S^z$ with $g^{zz} \simeq 7.6$ ( $g^{xx}{\simeq} g^{yy} \ll g^{zz}$ ). }

\begin{figure}
  \centering
  \includegraphics[width=0.95\columnwidth]{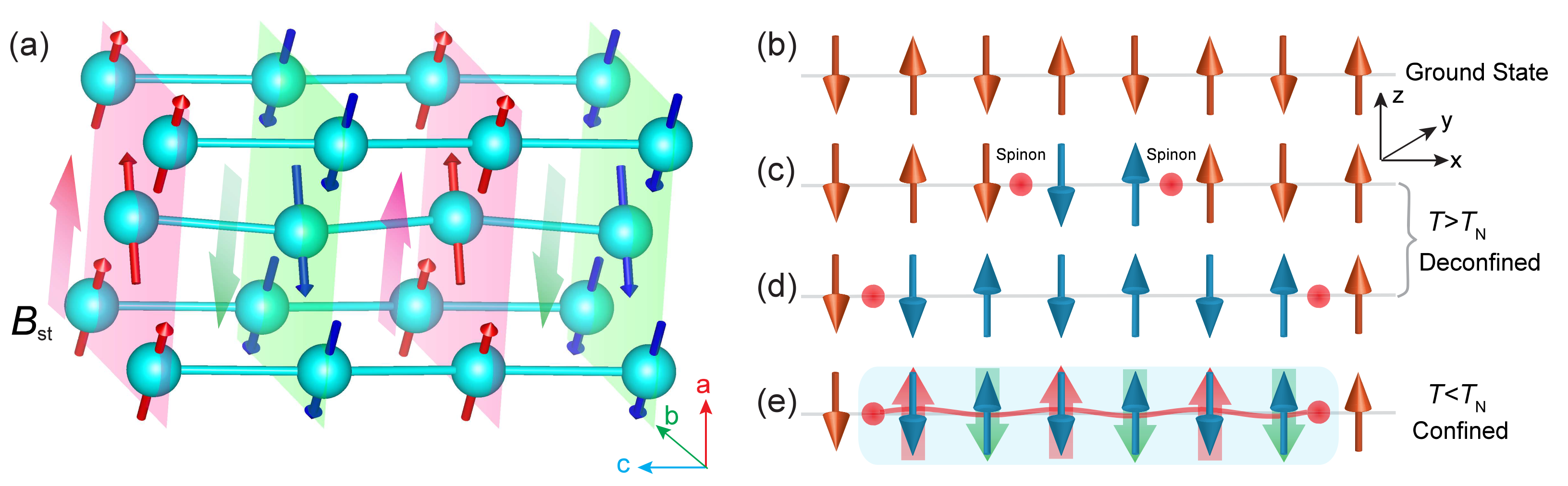}
  \caption{ Illustration of the magnetic structure and spinon confinement in YbAlO$_3$. (a) Magnetic structure below $T_{\rm N}=0.88$~K, where the Yb Ising moments align antiferromagnetically (AFM) in chains along the $c$-axis. These one dimensional (1D) chains are ferromagnetically (FM) coupled in $ab$-plane, which results in a net staggered field ($B_{\rm st}$), as indicated by the large red and blue arrows. (b)-(e) Sketch of the spinon generation and confinement in a 1D AFM chain. Since only longitudinal fluctuations of $\Delta S^{z}=0$ are observed in YbAlO$_{3}$, even numbers of Yb moments are flipped in the AF ground state (b), creating pairs of spinons (c). These spinons can propagate freely (deconfined) along the chain at $T>T_{\rm N}$ (c-d), while in the presence of the effective staggered field at $T<T_{\rm N}$, propagation of the spinons costs energy (confined) as the separation increases (e).}
  \label{SpinonSketch}
\end{figure}

The magnetic coupling between different sublattices is dominated by the long-range dipole-dipole interaction, which results in a weak ferromagnetic (FM) Ising-like coupling of type $S_i^z S_j^z$ in terms of the effective spins (see Supplementary Note 2). In contrast, the intrachain interaction between neighboring spins is dominated by antiferromagnetic (AFM) super-exchange. This combination leads to an $AxGy$-type AFM ordering~\cite{Radhakrishna1981} below the N\'eel temperature $T_{\rm N}=0.88$~K (see Fig.~\ref{SpinonSketch}a). Above $T_{\rm N}$, the weaker dipole-dipole interchain interaction can be neglected, since thermal fluctuations have destroyed the interchain ordering. In this free-standing 1D chain picture, the elementary excitations are fractional spinons, instead of the conventional magnons ($S=1$) of the magnetically ordered state. As illustrated in Fig.~\ref{SpinonSketch}b-d, a pair of domain walls can propagate along the chain direction, behaving as  free spin 1/2 spinons  for $T>T_{\rm N}$. Below $T_{\rm N}$, the  staggered interchain molecular field ($B_{\rm st}$) produces a confining potential that increases linearly in the distance between both spinons (Fig.~\ref{SpinonSketch}e). Consequently, spinons get confined into bound states at temperatures below $T_{\rm N}$~\cite{Lake2010}.

\begin{figure}
  \centering
  \includegraphics[width=0.95\columnwidth]{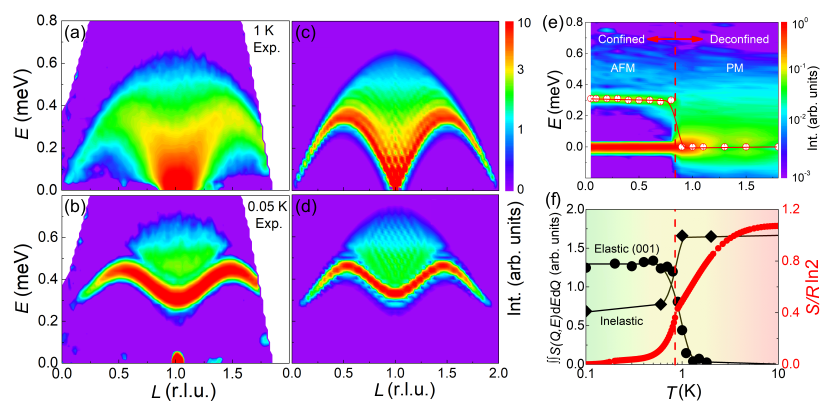}
  \caption{ Spinon confinement of YbAlO$_3$ in zero field. Inelastic neutron scattering spectra with background corrected measured at 1~K (a) and 0.05~K (b), compared to DMRG calculation with $B_{\rm st}/J=0$ (c), and $B_{\rm st}/J=0.27$ (d).  The DMRG calculation is performed on a 64-site periodic chain with bond dimension $M=400$ states at $T=0$, with intrachain interaction $J=0.21$~meV. In (d), the elastic peak at $L=1$ has been removed by hand. (e) Temperature evolution of the energy dependent neutron scattering spectrum integrated in window $H=[-0.1, 0.1]$~r.l.u., $K=[-0.1, 0.1]$~r.l.u., $L=[0.9, 1.1]$~r.l.u. A gapless continuum is observed at high temperatures, whereas a gap of about 0.3~meV (red circles) opens just at $T_{\rm N}$, due to the appearance of the staggered field. (f) Temperature dependence of the elastic (001) magnetic peak intensity (black circles), and inelastic neutron spectrum (black diamonds) integrated in the window $H=[-0.1, 0.1]$~r.l.u., $K=[-0.1, 0.1]$~r.l.u., $L=[0, 2.0]$~r.l.u., and $E=[0.1, 0.6]$~meV. The elastic and inelastic scatterings correspond to the statically ordered and fluctuating magnetic moments, respectively. The red circles are the temperature dependent entropy ($S$). The solid lines are a guide to the eyes, and the vertical red dashed line indicates the spinon confinement-deconfinement transition at $T_{\rm N}$.}
  \label{B_0T}
\end{figure}

In view of these considerations, the magnetic properties of YbAlO$_3$ can be described with an effective 1D spin-1/2 Hamiltonian
\begin{equation}\label{Hamiltonian}
\mathcal{H}=J\sum_i \bm{S}_i \cdot \bm{S}_{i+1}-B_{\rm st}\sum_{i}(-1)^{i}S_{i}^{z}-B_{\rm ex}\sum_{i}S_{i}^{z},
\end{equation}
where $x, y$ and $z$ are defined in Fig.~\ref{SpinonSketch}b, $J$ is the intrachain Heisenberg exchange, $B_{\rm st}$ is the effective staggered molecular field generated below $T_{\rm N}$, and $B_{\rm ex}$ is the external magnetic field applied along the local moments easy $z$-axis.  The pseudospin $S=1/2$ operators $S^a$ act on the Kramers doublet, consisting of the $m_J =\pm 7/2$ states to a good approximation,  which explains the $g$-tensor anisotropy. In other words,  the magnetic field and the magnetic moment of the neutron can couple only to the diagonal operator $S^z$ and the INS spectrum of YbAlO$_3$ is dominated by longitudinal fluctuations: the very weak  g-factor along the transverse  directions, $(g^{xx}/g^{zz})^2 \approx 1/273$, renders transverse fluctuations, $S^{xy} (\bm{Q},E)$, `hidden' to neutrons. Thus, the magnetic excitations observed in YbAlO$_3$ only arise from longitudinal fluctuations, as demonstrated by the neutron polarization factor presented in Supplementary Note 4, and by all the simulated spectra shown in Fig.~\ref{B_0T}c-d. As we will see later, the exchange anisotropy turns out to be surprisingly small in this material. This unexpected situation encountered in  an increasing number of compounds~\cite{Wu2016,Rau18,Hester18}, has been analyzed theoretically in \cite{Wu2016} and later in greater  detail  in Ref.~\cite{Rau18}.

Single crystal inelastic neutron scattering measurements have been performed to match the Hamiltonian parameters to experimental observables. The zero-field spectra ($B_{\rm ex}=0$) are presented in Fig.~\ref{B_0T}a-b at different temperatures (see also  Supplementary Note~3). Gapless spinon excitations along the chain direction ($00L$) are observed at 1.0 K ($T>T_{\rm N}$), with a broad two-spinon continuum extending from 0 up to about 0.7~meV in energy transfer (Fig.~\ref{B_0T}a).
No gap in the spin excitation can be resolved in the {paramagnetic (PM)} state of YbAlO$_3$ at the magnetic Brillouin zone center of $\bm{Q}$=(001), within the instrumental resolution about 0.05~meV.
Note that the temperature (1.0~K) is comparable to the instrumental resolution (0.05~meV), both of which are much smaller than the bandwidth of the two-spinon continuum in Fig.~\ref{B_0T}a, implying that  we can safely exclude a scenario where the continuum arises  from thermal or instrumental broadening (see also Supplementary Note 5).

In contrast, and in agreement with the analytical result for the spin-1/2 Heisenberg model in a staggered magnetic field \cite{Essler97}, a significant gap is observed below the ordering temperature, with a massive triplet mode and multi-spinon continuum extending from about 0.3 to 0.7~meV (Fig.~\ref{B_0T}b).

As demonstrated in Ref.~\onlinecite{Essler97}, the staggered field leads to  strong confinement of the spinons. In the continuum limit, one can use bosonization and the Hamiltonian is reduced to the sine-Gordon model. The corresponding Lagrangian density is
\begin{equation}
{\cal L} = \frac{1}{2}\Big[c^{-1}(\partial_{\tau}\phi)^2 + c(\partial_x\phi)^2\Big] - \alpha B_{\rm st}\sin(\beta\phi), ~~ \beta^2 = 2\pi, ~~ c = \pi J/2. \label{sG}
\end{equation}
and the spin is related to field $\phi(x)$:
\begin{equation}
S^z_n = \frac{\beta}{2\pi}\partial_x\phi + (-1)^n A\sin(\beta\phi), \label{Sz}
\end{equation}
where $A$ is a nonuniversal  amplitude. Here the staggered field is determined self-consistently as the one generated by the ordered moments of neighboring chains: $B_{\rm st} = J_{\perp}\langle S^z_{\rm st}\rangle$. The exact solution of the sine-Gordon model for this value of $\beta$ yields the spectra $E_n(p) = \sqrt{(cp)^2 + \Delta_n^2}$ with two excitation branches corresponding to a triplet $n=1$ and singlet $n=2$ with $\Delta_2 = \sqrt 3\Delta_1, ~ \Delta_n \sim (J B_{\rm st}^2)^{1/3}$. The inverse gap $c/\Delta \sim (J/B_{\rm st})^{2/3}$ is the confinement radius.  Since $B_{\rm st}$ is proportional to the order parameter, it vanishes at the transition point and the confinement radius becomes infinite. The triplet excitations with $S=1$ are seen in the dynamical  spin susceptibility.

The simulated longitudinal spin excitation spectrum is presented in Fig.~\ref{B_0T}c-d, using 64-site density matrix renormalization group calculations (DMRG) at $T=0$ with periodic boundary conditions~\cite{White1992,White1993}. The Hamiltonian (\ref{Hamiltonian}) captures the main features of the experimental data measured at 1~K and 0.05~K for intrachain exchange $J=0.21$~meV, staggered field $B_{\rm st}$=0 (Fig.~\ref{B_0T}c) and $B_{\rm st}/J=0.27$ (Fig.~\ref{B_0T}d). It is worth noting that a finite temperature DMRG calculation  for the same model~\cite{Lake13} can account for  the deviations between the experimental data and our DMRG result at $T=0$.

The nearly isotropic nature of the intrachain interaction may look  surprising if we consider that the Yb$^{3+}$ ground state doublet is highly anisotropic. However, as was pointed out in Ref.~\onlinecite{Wu2016}, the exchange interaction of rare earth ions has a form of permutation operator which has matrix elements between all states. Being projected on the lowest Kramers doublet by the crystal field, it acquires the familiar ${\bf S}_i \cdot {\bf S}_{i+1}$ form. Apart from Yb$_2$Pt$_2$Pb and YbAlO$_3$ the only known previous example of Heisenberg-like Yb chains is Yb$_4$As$_3$~\cite{Kohgi1997,Kohgi2001}. It should be noted, however, that the shape of the neutron scattering spectra is not sufficient to determine the magnitude of a possible eaxy-plane exchange anisotropy $\Delta$ that would still be compatible with the observed TLL at zero magnetic field: for an XXZ chain, the transverse (longitudinal) modes for $0.25\lesssim \Delta \lesssim 1$ ($0.5\lesssim \Delta \lesssim 1$) almost do not change~\cite{Caux2005,Caux2012}. Nevertheless, as we explain in a later  section, the observation of a free fermion fixed point at the saturation field $B=B_{\rm s}$ excludes the possibility of a significant  easy-plane exchange anisotropy.

The temperature evolution of the energy dependent scattering at the AFM zone center $\bm{Q}$=(001) is presented in Fig.~\ref{B_0T}e. The continuous temperature dependent entropy $S$ in Fig.~\ref{B_0T}f, suggests a second order transition occurs at $T_{\rm N}$. The gap in spin excitations closes within a narrow temperature $T_{\rm N}\pm0.1$~K, which is unlikely due to thermal broadening. Figure~\ref{B_0T}f shows the temperature dependence of the (001) Bragg peak intensity, indicating that static moments build up at the phase transition. Meanwhile, the integrated inelastic neutron intensity shows that about half of the total  spectral weight remains inelastic below $T_{\rm N}$ (see Fig.~\ref{B_0T}f). The temperature dependence of entropy ($S$) is over-plotted in Fig.~\ref{B_0T}f as well. As expected from the ground state doublets, the full {entropy ($ R \cdot\rm ln2$)} is reached above $T=10$~K, however, only about $0.4 R\cdot \rm ln2$ is released at $T_{\rm N}$. This value is consistent with the  neutron scattering results: about half of the magnetic moment keeps fluctuating in the ordered state.

\subsection*{Magnetic field-induced quantum phase transitions}

\begin{figure}
  \centering
  \includegraphics[width=0.95\columnwidth]{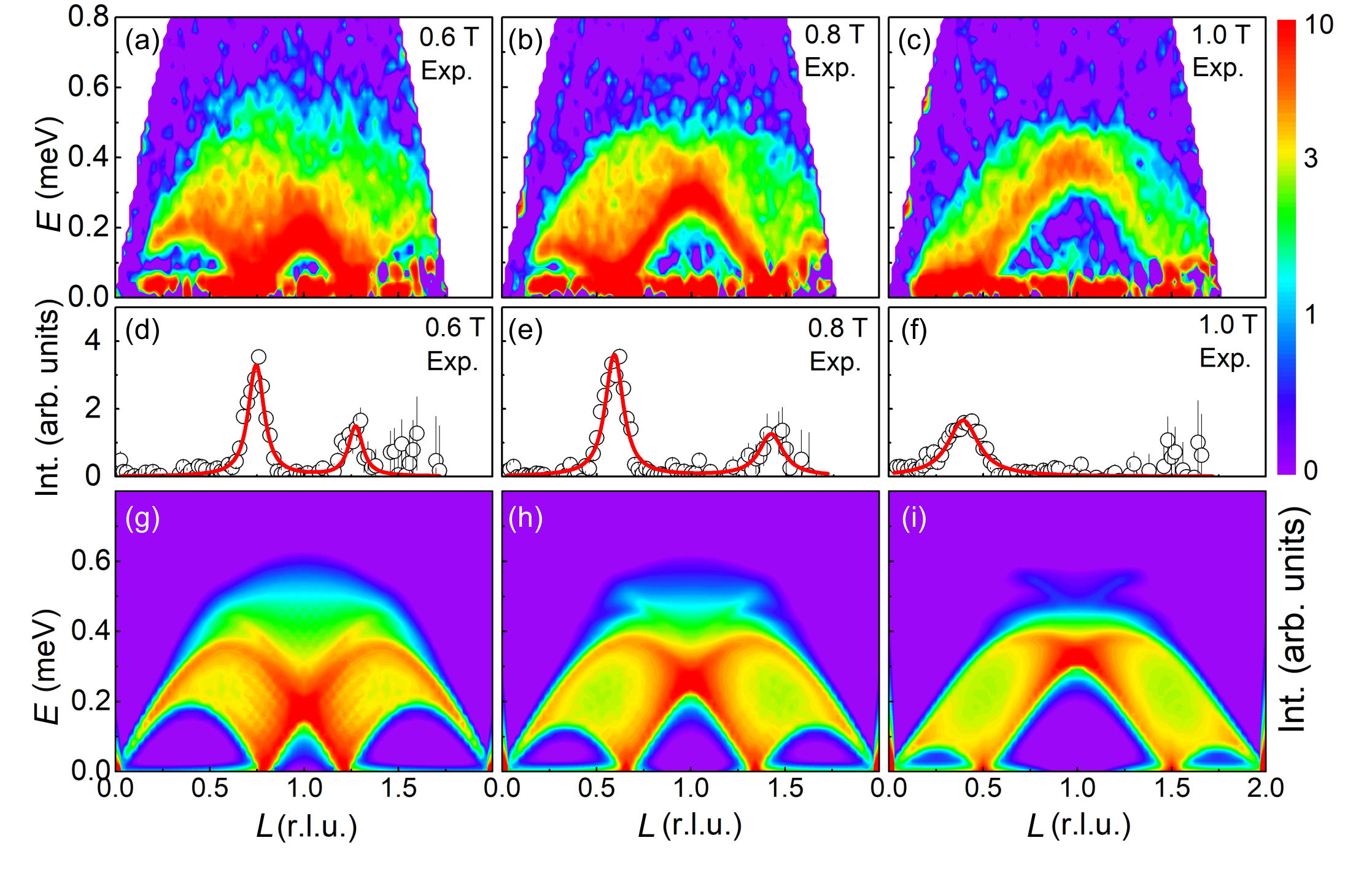}
  \caption{Magnetic field tuned spin excitation of  YbAlO$_3$. (a)-(c) Inelastic neutron scattering spectra with background corrected measured at 0.05~K, in different magnetic fields applied along the $a$-axis, of 0.6~T (a), 0.8~T (b), and 1.0~T (c). (d)-(f) The $Q$-cuts of in-field experimental data along wave vector ($00L$), integrated over $H=[-0.1, 0.1]$~r.l.u., $K=[-0.6, 0.2]$~r.l.u., and $E=[0.05, 0.06]$~meV. The peak positions indicate the zero energy soft modes, where the gap closes. In the raw data, the error bar is equal to the square root of the number of counts (Poisson statistics). In the processed data, where we have various efficiency corrections and we need to account for different measurement statistics at each point, we used the rules for propagation of errors of uncorrelated measurements. (g)-(i) DMRG simulated spin excitation spectrum, with staggered field $B_{\rm st}=0$, and external field $B_{\rm ex}/J=0.8$ (g), $B_{\rm ex}/J=1.2$ (h), and $B_{\rm ex}/J=1.6$ (i).}
  \label{B_fields}
\end{figure}

\begin{figure}
  \centering
  \includegraphics[width=0.8\columnwidth]{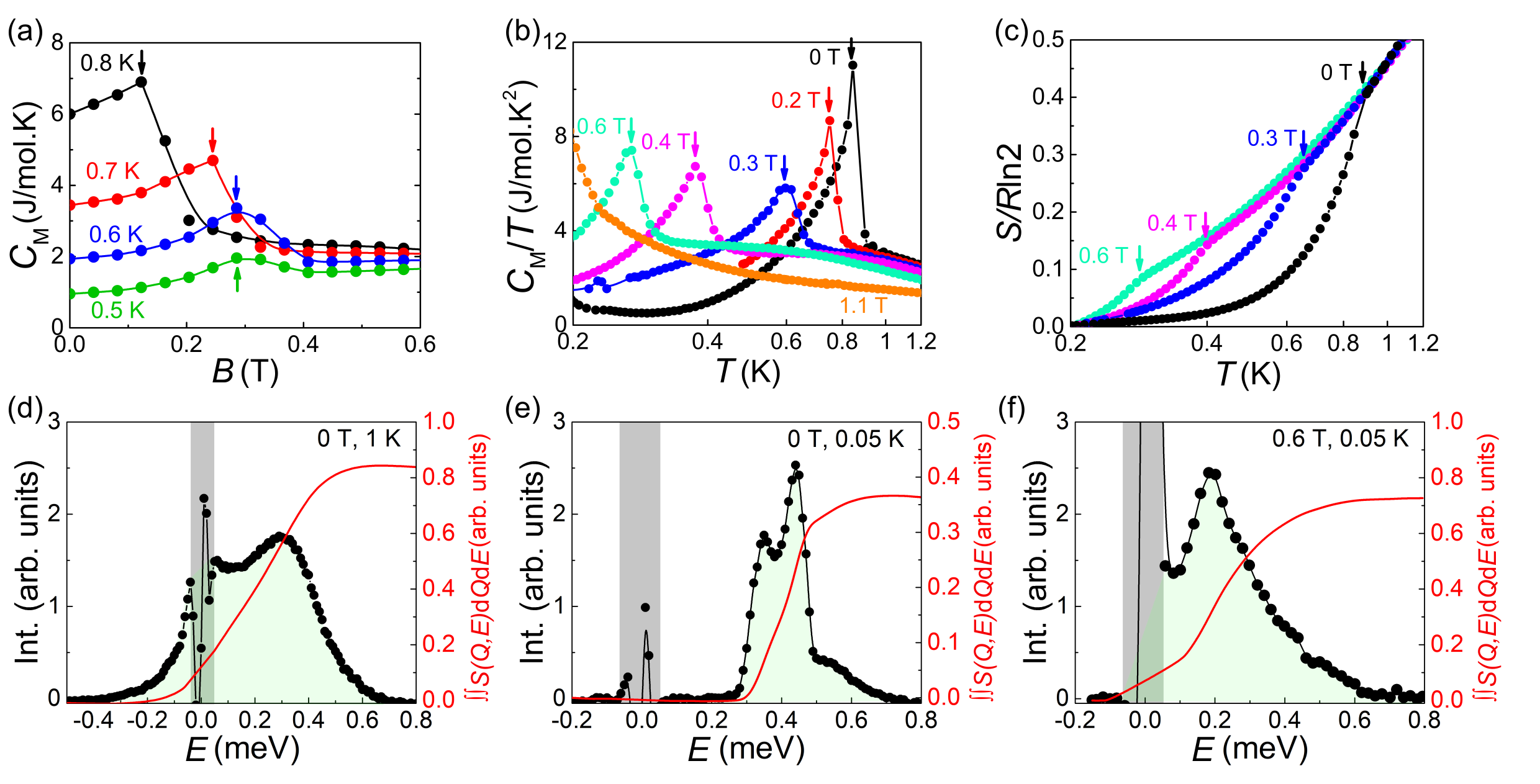}
  \caption{Fluctuating moments in fields. (a) Field dependent magnetic specific heat $C_{\rm{M}}/T$ at different temperatures, as indicated. (b) Temperature dependent magnetic specific heat $C_{\rm{M}}/T$ in different fields. (c) Temperature dependent entropy, normalized to $R\cdot\ln{2}$, assuming doublet ground states.
  The arrows indicate the magnetic phase transition. (d)-(f) Energy dependent neutron scattering intensity, integrated over wave vector $H=[-0.2,0.2]$~r.l.u., $K=[-1,1]$~r.l.u., and $L=[0,2]$~r.l.u., at different temperatures and fields, as indicated.
  The red lines are the integrated intensity over the energy spectrum (green shadow area) up to 0.8~meV. The gray bars indicate the instrumental resolution of about $\pm0.05$~meV.}
  \label{Deconfinement}
\end{figure}

\begin{figure}
  \centering
  \includegraphics[width=0.95\columnwidth]{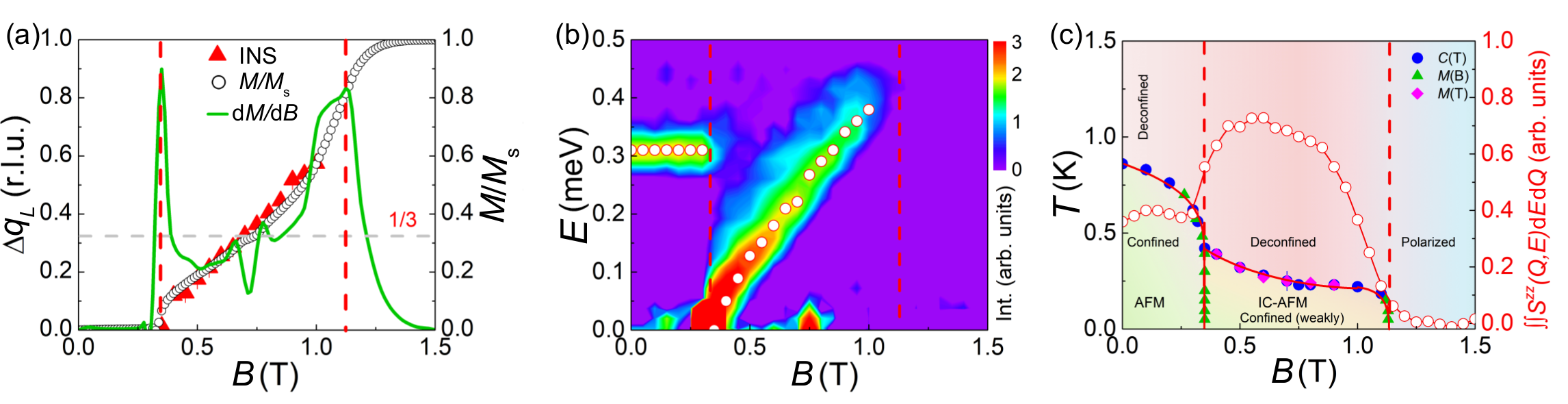}
  \caption{Field dependent magnetization and magnetic phase diagram of YbAlO$_3$. (a) Field dependent zero energy soft point at wave vector shift $\Delta q_{\rm L}=Q_{\rm L}-1$, extracted from the INS. Normalized magnetization $M/M_{\rm s}$, measured at 0.05~K, and the corresponding magnetic susceptibility ${\rm d}M/{\rm d}B$ are over plotted. $M_{\rm s}$ is the saturation moment for fields along the $a$-axis, and the horizontal dashed lines indicate the 1/3 magnetization plateau. (b) Field evolution of the energy dependent inelastic neutron spectrum, at 0.05~K, integrated over wave vector $H=[-0.1, 0.1]$~r.l.u., $K=[-0.6, 0.2]$~r.l.u., $L=[0.9, 1.1]$~r.l.u.. The empty circles indicate the peak positions. (c) The field temperature ($B-T$) magnetic phase diagram. The field dependent INS spectrum measured at 0.05~K is over plotted, which is integrated over $H=[-0.1, 0.1]$~r.l.u., $K=[-0.6, 0.2]$~r.l.u., $L=[0, 2]$~r.l.u. and $E = [0.05, 0.6]$~meV. The vertical dashed lines indicate the critical field $B_{\rm c}=0.35$~T, and saturation field $B_{\rm s}=1.13$~T, defined through the peak positions of ${\rm d}M/{\rm d}B$. }
  \label{Neutron_phase_diagram}
\end{figure}

With a magnetic field applied along the a-axis, the static AFM order is suppressed at the critical field $B_{\rm c}=0.35$ T in favor of incommensurate magnetic ordering. As entering the IC-AFM phase at $B_{\rm c}$, the spin excitation gap in the magnetic zone center $\bm{Q}=\rm (001)$ closes abruptly (Fig.~\ref{B_fields}a-c,; Fig.~\ref{Deconfinement}e,f and Fig.~\ref{Neutron_phase_diagram}b).
Figure~\ref{Deconfinement}a shows the field dependence of the magnetic contribution to the specific heat $C_{\rm{M}}/T$ around $B_{\rm{c}}$. The step-like anomaly becomes smaller and broader upon decreasing $T$ and finally evolves into a weak maximum around $B_{\rm{c}}$ (see also Supplementary Note 1). Meanwhile, the field temperature ($B$-$T$) phase line becomes very steep (${\rm d}B/{\rm d}T\rightarrow-\infty$) as $T$ approaches 0~K (Fig.~\ref{Neutron_phase_diagram}c). This indicates that the system is tuned through a first order phase transition at $T=0$~K~\cite{Flouquet2005}, which is consistent with the neutron scattering observations (Fig.~\ref{Neutron_phase_diagram}b).

For $B>B_{\rm{c}}$, the longitudinal ($S^{zz} (\bm{Q},E)$) and transverse ($S^{xx} (\bm{Q},E)$, $S^{yy} (\bm{Q},E)$) components of the dynamical spin structure factor behave differently~\cite{Muller1981}. The bosonization formula (\ref{Sz}) is modified in the presence of finite magnetization $\langle S^z\rangle = M$ :
\begin{equation}
S^z_n = M + \frac{\beta}{2\pi}\partial_x\phi + A\sin(\beta\phi + 2\pi Mn),
\end{equation}
Therefore, in the field-induced incommensurate phase, the interchain interaction becomes
\begin{equation}
\sum _{i,j}J_{ij}S^z_iS^z_j \rightarrow \sum_{i,j}J_{ij}A^2\cos[\beta(\phi_i - \phi_j)]. \label{V}
\end{equation}
The oscillatory term does not survive after summing over the phases.
As we have mentioned above, the spectral gap is suppressed when the magnetic field exceeds the critical value, and the single-chain system becomes critical.
However, the interchain coupling remains relevant and leading to long-range magnetic ordering and the corresponding thermodynamic phase transition (see Fig.~\ref{Deconfinement}b and Fig.~\ref{Neutron_phase_diagram}c).
Below the transition one can expand the cosine in (\ref{V}) and obtain the quadratic action for $\phi$ so that the interaction leads to a gapless phason mode, which explains a finite low energy spectral weight on Fig.~\ref{Deconfinement}f, and a significant $C_{\rm M}/T$ values below the transition:
\begin{equation}
\omega^2 = v^2q_{\parallel}^2 + \sum_ja_j\sin^2({\mathbf{q e}}_j/2).
\end{equation}
At higher energies one can again approximate $\sum_j\cos[\beta(\phi_i - \phi_j)] \approx \cos\beta\phi_i\sum_j\langle \cos\beta\phi_j \rangle$
to obtain the effective sine-Gordon model, as it was done in Ref.~\onlinecite{Essler97}. The gapped $S=1$ excitations of the sine-Gordon model are allowed to decay into low energy phasons and become resonances.

Figure~\ref{B_fields}a-c and Fig.~\ref{Neutron_phase_diagram}b include the experimental spin excitation spectra of YbAlO$_3$ for different external fields. The low energy features discussed above are obscured in the color pictures since the main spectral weight remains in the single chains.
The $\bm{Q}$-cuts show that the incommensurate (IC) zero energy soft modes are symmetrically located around the AF zone center (Fig.~\ref{B_fields}d-f)~\cite{Muller1981}. The shift of these IC wave vectors $Q_L=1\pm\Delta q_L$ is directly proportional to the field-induced magnetization $M$, i.e., $\Delta q_L\sim 2\pi M/M_s$. As one can see in Fig.~\ref{Neutron_phase_diagram}a, the INS zero energy soft point traced the magnetization curve up to about 1.0 T. Finally, all the magnetic moments become fully polarized  above the saturation field $B_{\rm{s}}=1.1$~T. As we have emphasized before, no magnon-like spectra can be observed for $B>B_{\rm{s}}$, because the modes become purely transverse.

For $B_{\rm{c}}<B<B_{\rm s}$, the Yb$^{3+}$ moments are highly fluctuating even in the ordered phase. Figure~\ref{Deconfinement}d-f shows the energy dependent neutron scattering intensity integrated over the first Brillouin zone. The integrated intensity up to 0.8~meV, $S_{\rm{int}}=\iint{S}\left(\bm{Q},E\right){\;}{\rm d}\bm{Q}{\rm d}E$, is then an estimate of the square of the fluctuating moment. In zero field, $S_{\rm{int}}$ at 0.05~K (AFM ordered state) is about $0.37/0.85\simeq44\%$ (normalized by the intensity at $T=1$ K and $B=0$); while in 0.6~T, it reaches about $0.72/0.85\simeq85\%$. Similar conclusions are found in the temperature dependent specific heat and entropy as well. The integrated temperature dependent entropy shown in {Fig.~\ref{Deconfinement}c}, reveals that only $10\sim20\%$ of $R\cdot\ln2$ is released at the IC-AFM phase transition.
These together suggest that large moment fluctuations coexist with a small ordered moment ($10\sim20\%$ of the full moment) in the IC-AFM phase~(Fig.~\ref{Neutron_phase_diagram}c).

Since the interchain molecular field is proportional to the magnitude of the ordered moments, in a first approximation we can ignore it for $B_{\rm c}<B<B_{\rm s}$. This simplification reduces the Hamiltonian (1) to the conventional Heisenberg model in an external field $B_{\rm ex}$. We again use the DMRG calculation to obtain the longitudinal spin structure factor $S^{zz} (\bm{Q},E)$ above $B_{\rm c}$, in various external fields $B_{\rm ex}/J=0.8,1.2,1.6$, as presented in Fig.~\ref{B_fields}g,h,i. We note that, although a self-consistent treatment of the static moments from neighboring chains is required to obtain a more accurate excitation spectrum, overall qualitative agreement with the experimental data is already achieved at this level of approximation.

It is interesting to note that  an $M/M_{\rm s}=1/3$ magnetization plateau appears at 0.75~T, indicating the stabilization of a new magnetic structure. A rather abrupt increase of the magnetic susceptibility is also observed at $M/M_{\rm s}=1/2$, although no obvious magnetization plateau can be resolved.

A key and surprising observation of this work is that YbAlO$_3$ can be modeled by nearly isotropic Heisenberg spin 1/2 chains, despite the very strong $g$-tensor anisotropy. This unusual combination together with the dominant dipolar interchain interactions make YbAlO$_3$ an ideal material
for studying the competition between sliding TLL's and spin density wave ordering. Our measured phase diagram reflects such competition. Each chain
behaves as a TLL at temperatures $T_{\rm N}<T\ll J$, while spin density wave ordering (SDW) develops below $T_{\rm N}$. The intrachain Hamiltonian has a divergent longitudinal magnetic susceptibility $\chi(q,\omega)$ at $q=2k_{\rm f}$ and $\omega=0$, where $k_{\rm f}$  is the Fermi wave-vector of the fermionic system associated with the TLL. Given that $k_{\rm f}=\pm (m/2+1/2)\pi$, with $m=M/M_{\rm s}$, $k_{\rm f}$  increases linearly with $M$ between $B_{\rm c}$ and $B_{\rm s}$ ($k_{\rm f}=\pi/2$  for $0\leq B<B_{\rm c}$). The Ising-like interchain dipolar interaction favors a collinear structure with maximum amplitude of the local moment at each site. However, the magnitude of the moment $|\langle S_i^{z} \rangle|$  is necessarily modulated for general values of $2k_{\rm f}$, leading to competition between intrachain exchange and the interchain Ising-like dipolar interaction that penalizes any longitudinal modulation  of the magnetic moments.
The only exceptions are $2k_{\rm f}$ equal to 0 ($\uparrow\uparrow\uparrow$),$\pi$ ($\uparrow\downarrow\uparrow\downarrow$) or $\pm2\pi/3$($\uparrow\uparrow\downarrow$). The magnitude of the local moments is not modulated for these particular ordering wave vectors, implying that intrachain and interchain interactions can be satisfied simultaneously.
Note that the $\uparrow\uparrow\downarrow$ ordering only contains Fourier components $q=\pm2\pi/3$ plus the field-induced $q=0$ component. The lack of competition for these special cases explains the observed magnetization plateaus at $m=0,1$ and 1/3. A spin density wave with dominant $2k_{\rm f}$ ordering wave-vector is expected for other magnetization values, as it is confirmed by the fact that the ordering wave vector extracted from the INS data tracks the magnetization curve (see Fig.~4a). However, we expect that higher harmonics will be induced for this magnetization values producing spin superstructures similar to those emerging from  the anisotropic next-nearest-neighbor Ising (ANNNI) model~\cite{Elliott1961,Fisher1980,Selke1988}.

\subsection*{Quantum critical scaling and Universality class}

The quantum phase transition at the saturation field is of second order, implying a quantum critical point (QCP) at $B=B_s$ and $T=0$. This QCP is a  Gaussian fixed point because the effective dimension of the low-energy effective theory is $D=3+1$ (the dynamical exponent is $z=1$ because of the exchange anisotropy). For temperatures higher than the exchange anisotropy and the interchain coupling, we expect a crossover into a regime controlled by a free fermion fixed point ($\nu=1/2$) in dimension $D=1+2$ (the dynamical exponent becomes $z=2$ because the field couples to a conserved quantity in absence of exchange anisotropy)~\cite{Sachdev1999,Zapf2014}.

\begin{figure}
  \centering
  \includegraphics[width=0.95\columnwidth]{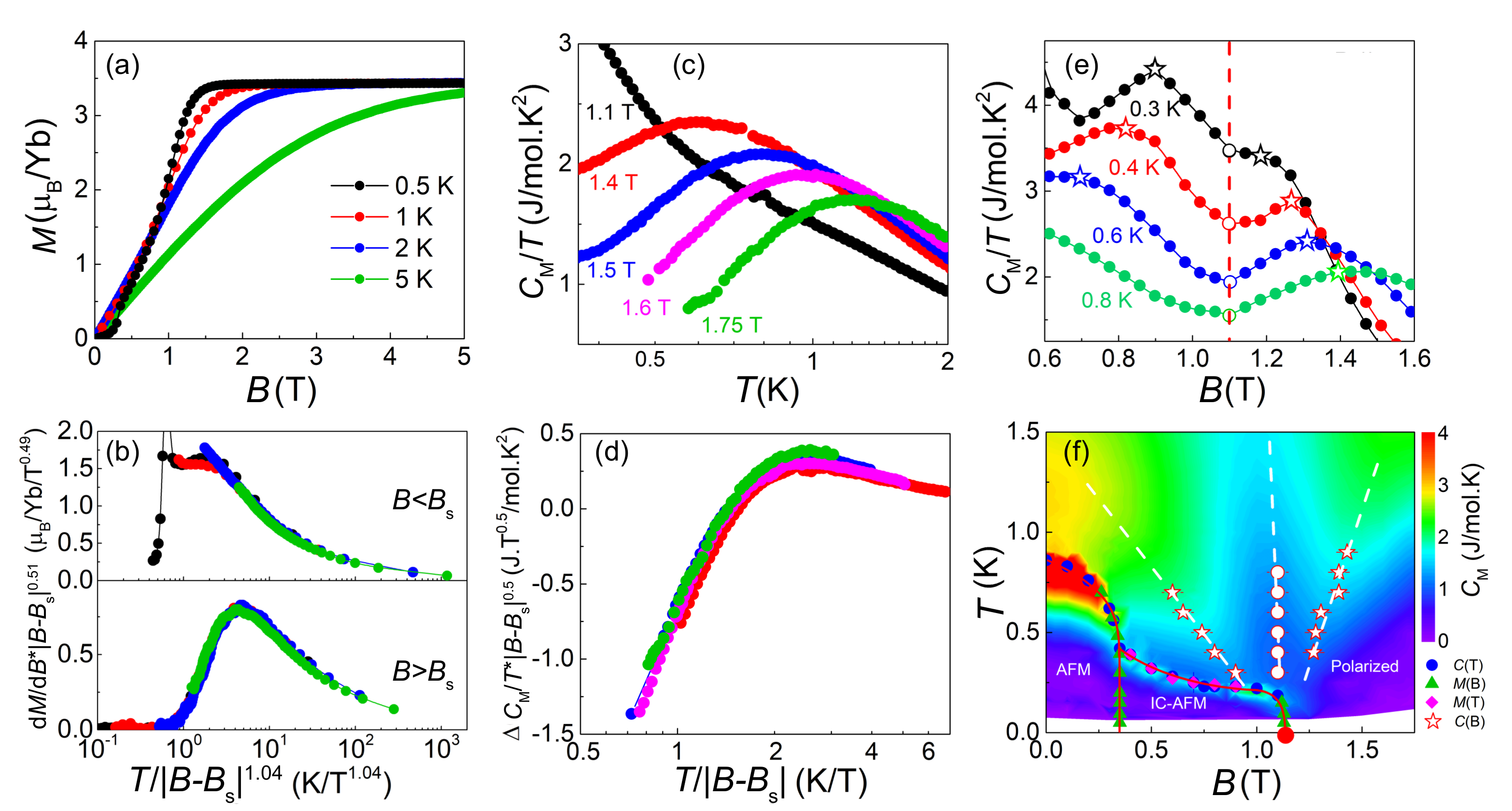}
  \caption{Quantum critical scaling and universality of YbAlO$_3$. (a) Field dependent magnetization $M$ measured at different temperatures. (b) Critical scaling of the magnetic susceptibility in (a), for fields both below and above $B_{\rm s}$.
  (c) Temperature dependent magnetic specific heat $C_{\rm M}/T$ measured at different fields. (d) Critical scaling of the field dependent magnetic specific heat $\Delta C_{\rm M}/T$ shown in (c).
  (e) Field dependent magnetic specific heat $C_{\rm M}/T$ at different temperatures. The red dashed line indicates the saturation field $B_{\rm s}=1.1$~T at the minimum, and crossovers with two broad maximum shoulders are observed. (f) Contour plot of the magnetic specific heat $C_{\rm M}$ as function of temperature and field. Magnetic phase boundaries extracted through different measurements are over-plotted. The red circles indicate the saturation field $B_{\rm s}$ at different temperatures, and the empty stars indicate the crossover defined through the field dependent specific heat maxima.}
  \label{QCPScaling}
\end{figure}

Figure~\ref{QCPScaling}a shows the field dependent magnetization $M$ at different temperatures for $T > 0.5$ K. Near the QCP, the magnetic susceptibility scales as
\begin{equation}\label{Scaling2}
 \frac{{\rm d}M}{{\rm d}B}=\tilde{B}^{\nu(d+z)-2}\varphi(\frac{T}{\tilde{B}^{\nu z}}),
\end{equation}
where $d$ is the spatial dimension and $\tilde{B}=|B-B_{\rm s}|$.
The values $d$, $\nu$ and $z$ can be obtained by minimizing the deviations from this scaling behavior (see Supplementary Note 6), giving:
\begin{gather}\label{vz}
 \nu z=1.04,\\
 2-\nu(d+z)=0.51.
\end{gather}
As shown in Fig.~\ref{QCPScaling}b, this set of exponents collapses the measured susceptibility data.

In addition, the field dependent specific heat should obey the following scaling relation~\cite{Schroder2000,Matsumoto2011,Bianchi2003,Wu2014}:
\begin{equation}\label{Scaling}
 \frac{\Delta C_{\rm M}}{T}= \tilde{B}^{\nu(d-z)}\psi(\frac{T}{\tilde{B}^{\nu z}})=\tilde{B}^{-0.5}\psi(\frac{T}{\tilde{B}}),
\end{equation}
where
\begin{equation}
\frac{\Delta C_{\rm M}}{T}=\frac{C_{\rm M}(B)}{T}-\frac{C_{\rm M}(B_{\rm s})}{T}.
\end{equation}
The measured temperature dependent $C_{\rm M}/T$ curves, shown in Fig.~\ref{QCPScaling}(c),
collapse into a single curve for $T\geq 0.3$~K, once they are rescaled with $\nu z=1$ and $d=1$ (see Fig.~\ref{QCPScaling}d).

$C_M(B)/T$ exhibits a clear two-peak structure in the vicinity of the quantum critical region for each fixed temperature $T\geq0.3$~K (star markers in Fig.~\ref{QCPScaling}e). This is consistent with the expectation of the 1D TLL behavior as discussed in Ref.~\onlinecite{Blanc2018}. Defining this peak position as the crossover temperature $T^*$ into the quantum critical regime, the linear dependence $T^{*}\sim \tilde{B}^{\nu z}$ with $\nu z=1$ for $T\geq0.3$~K again agrees with the free fermion fix point (Fig.~\ref{QCPScaling}e,f).
We note that the free fermion fixed point is only compatible with an isotropic exchange interaction because the applied magnetic field is nearly orthogonal to the uniaxial symmetry chain axis (as shown in Fig.~\ref{SpinonSketch}, the magnetic moments are nearly perpendicular to the chains). Any type of exchange anisotropy would then  change the universality class of this QCP to Ising in D=1+1 ($z=\nu=1$). The single-chain Hamiltonian must then be isotropic (Heisenberg) for the model the retain a U(1) symmetry in the presence of the external magnetic field. Based on the above-described analysis and the value of the exchange parameter ($J\approx 0.21$~meV), we can narrow down the range of the exchange anisotropy to $0.88\lesssim \Delta \lesssim 1$.

\section*{Discussion}

Our results demonstrate that $f$-electron based magnets can provide realizations of various aspects of quasi-one dimensional physics. This is so despite the presence of strong spin-orbit coupling combined with crystal field splitting that produce a ground state Kramers doublet with well separated values of $m_{\rm J}$. Naively one would think about such doublets should behave a as classical (Ising) spins. This unexpected situation,  which is appearing in an increasing number of Yb-based compounds~\cite{Wu2016,Rau18,Hester18}, enables the study of model Hamiltonians, which have traditionally been regarded as ``toy models'' due to the combination of interactions that do not appear together. The array of TLL's coupled by density-density interactions is a clear example of a model that was originally introduced to study the possible existence of sliding TLL phases, but it is difficult to find in real materials due to the very unusual combination of intrachain XXZ exchange  and pure Ising interchain interaction. YbAlO$_3$ provides a natural realization of the spin 1/2 version of this model, enabling the observation of a quantum fractional spinon continuum above $T_{\rm N}$ and the transition into a confined (magnetically ordered) low-temperature state characterized by massive  excitations. Both, the gapless spinon spectrum and the critical behavior around the saturation field at $T>T_N$   unambiguously demonstrate that the intrachain exchange is Heisenberg-like to a very good approximation.

Rau and Gingras \cite{Rau18} derived an isotropic super-exchange interaction for Yb-based compounds with edge-sharing octahedra. While it is not yet clear that this result applies directly to our material, it shows that Yb-based compounds can support nearly isotropic super-exchange of the type that we are finding in YbAlO$_3$. Also, note that further experiments such as an electron spin resonance would help quantify the small anisotropies in this system.

Our observations in YbAlO$_3$ suggest that it is worth exploring other members of the rare earth perovskite family RMO$_3$. Compared to most of the $d$-electron based spin chains with critical fields of the order 10-100~T, YbAlO$_3$ can be saturated with a field of order 1~T owing to the much weaker exchange interaction $J$ and large value of the effective $g$-factor. In addition, the crystal structure naturally adapts to existing standard perovskite thin film substrates. These properties can be exploited in future spinon-based research~\cite{Hirobe2016} for material engineering under easily-accessible laboratory conditions.

\section*{Methods}

\subsection*{Sample preparation and Thermal property measurements}
YbAlO$_3$ single crystals were grown by the Czochralski technique~\cite{Buryy2010}. Magnetization measurements were carried out using three magnetometers for different temperature ranges: commercial vibration sample magnetometer Quantum Design Magnetic Property Measurement System (MPMS-VSM) for the high-temperature measurements $T=1.8$ -- 400~K, MPMS-3 with $^{3}$He insert for the temperatures $T=0.5$ -- 4~K, and high-resolution capacitive Faraday magnetometer~\cite{Faraday} for the low-temperature range $T=0.05$ -- 0.9~K.
The specific heat measurements were carried out using the relaxation time method, with a Quantum Design Physical Property  Measurements System (PPMS) down to temperatures of 0.36~K and a custom compensated heat-pulse calorimeter with dilution refrigerator insert at MPI CPfS~\cite{Wilhelm2004}.

\subsection*{Neutron scattering}
Neutron scattering measurements were performed at two fixed incident energy of 3.32~meV ($\lambda_{i}=4.97$~{\AA}) and 1.55~meV ($\lambda_{i}=7.26$~{\AA}), with the time-of-flight Cold Neutron Chopper Spectrometer (CNCS)~\cite{Ehlers2011,Ehlers2016}, at the Spallation Neutron Source (SNS) at Oak Ridge National Laboratory. A single crystal of YbAlO$_3$ of about 0.59~g was aligned in the ($0KL$) scattering plane, with magnetic field along the vertical [100] direction. A dilution refrigerator insert that can access temperatures as low as 0.05~K was used. The software packages \textsc{Dave}~\cite{Dave} and \textsc{MantidPlot}~\cite{Mantid} were used for data reduction and analysis.

We applied two different methods to remove the background near $|\mathbf{Q}|=0$ from the direct beam. For zero field data we used an empty can file. For measurements under different fields, we used the high field data at 2~T for  background subtraction (almost all the magnetic inelastic signal of the Yb Ising moments is suppressed  above the saturation field of 1.1~T).
As presented in Fig.~\ref{B_0T}, Fig.~\ref{B_fields} and Fig.~\ref{Deconfinement}, the background correction produces some uncertainty within the instrumental resolution of $\pm0.05$~meV, but it removes the instrumental background for all the other inelastic energy range.

\section*{Data Availability}

The datasets generated during the current study are available from the corresponding author on reasonable request.

\newpage

\newpage

\newpage

\newpage

\newpage

\clearpage

\newpage

\newpage

\clearpage
\newpage

\section*{Acknowledgements}
We would like to thank G. Loutts for the help with single crystal growth, C. Klausnitzer for the help with magnetization measurements, and I. Zaliznyak, F. Ronning, M. C. Aronson and E-J. Guo for useful discussions. This research used resources at the High Flux Isotope Reactor and Spallation Neutron Source, a DOE Office of Science User Facility operated by the Oak Ridge National Laboratory. Research supported in part by the Laboratory Directed Research and Development Program of Oak Ridge National Laboratory, managed by UT-Battelle, LLC, for the U.S. Department of Energy. This work is partly supported by the U.S. Department of Energy (DOE), Office of Science, Basic Energy Sciences (BES), Materials Science and Engineering Division. This work used resources of the Compute and Data Environment for Science (CADES) at the Oak Ridge National Laboratory, which is supported by the Office of Science of the U.S. Department of Energy under Contract No. DE-AC05-00OR22725. This research used resources of the Oak Ridge Leadership Computing Facility, which is a DOE Office of Science User Facility supported under Contract DE-AC05-00OR22725. L.V. acknowledges Ukrainian Ministry of Education and Sciences for partial support under project 'Feryt'. Z.W. and C.D.B. are supported by funding from the Lincoln Chair of Excellence in Physics and by the LANL Directed Research and Development program. S.E.N. acknowledges support from the International Max Planck Research School for Chemistry and Physics of Quantum Materials (IMPRS-CPQM). A.C. is partially supported by the U.S. Department of Energy, Office of Science, Basic Energy Sciences, Materials Sciences and Engineering Division. W.Z. is supported by U.S. DOE through the LANL LDRD Program. M.B. would like to thank the DFG for financial support from project BR 4110/1-1.

\section*{Author contributions}
These authors contributed equally: L. S. Wu, S. E. Nikitin and Z. Wang.

L.S.W and A.P. designed the experiments. S.E.N. and M.B. carried out the low temperature thermal property measurements. Z.W. and W.Z. did the DMRG calculation and A.M.T. and C.D.B. developed theoretical explanations. A.M.S. and D.A.T. performed the classical spin simulation for the paramagnon excitations. L.V. provided single crystals used in this study. L.S.W., A.P., G.E., M.F. and G.S. performed neutron scattering experiments. A.T.S. supported the neutron data analysis. M.D.L., A.D.C., G.E., C.D.B., A.M.T. and A.P. supervised and supported the study. All authors discussed the results and contributed to the writing of the manuscript.

\section*{Competing interests}
The authors declare no competing interests.

\section*{Corresponding authors}
Correspondence and requests for materials should be addressed to L. S. Wu and A. Podlesnyak .


\begin{thebibliography}{10}

\bibitem{Wu2016}
Wu, L. S. et al. Orbital-exchange and fractional quantum number excitations in an $f$-electron metal,  Yb$_2$Pt$_2$Pb. {\it Science \bf 352}, 1206-1210 (2016).

\bibitem{Classen2018}
Classen, L., Zaliznyak, I. $\&$ Tsvelik, A. M. Three-Dimensional Non-Fermi-Liquid Behavior from One-Dimensional Quantum Critical Local Moments. {\it Phys. Rev. Lett. \bf 120},  156404 (2018).

\bibitem{Kim2013}
Kim, M. S. $\&$ Aronson, M. C. Spin Liquids and Antiferromagnetic Order in the Shastry-Sutherland-Lattice Compound Yb$_2$Pt$_2$Pb. {\it Phys. Rev. Lett. \bf 110}, 017201 (2013).

\bibitem{Ochiai2011}
Ochiai, A. et al. Field-Induced Partially Disordered State in Yb$_2$Pt$_2$Pb. {\it J. Phys. Soc. Jpn. \bf 80}, 123705 (2011).

\bibitem{Radhakrishna1981}
Radhakrishna, P., Hammann, J., Ocio, M., Pari, P. $\&$ Allain, Y. Antiferromagnetic ordering in the ytterbium aluminum perovskite YbAlO$_3$.  {\it Solid State Commun. \bf 37},  813 (1981).


\bibitem{Tomonaga1950}
Tomonaga, S., Remarks on Bloch's Method of Sound Waves applied to Many-Fermion Problems,  {\it Progr. Theoret. Phys. \bf 5}, 544 (1950)

\bibitem{Luttinger1963}
Luttinger, J. M., An Exactly Soluble Model of a Many-Fermion System, {\it J. Math. Phys. \bf 4}, 1154 (1963)

\bibitem{Mattis1965}
Mattis, D. C., $\&$ Lieb, E. H. Exact Solution of a Many Fermion System and Its Associated Boson Field, {\it J. Math. Phys. \bf 6}, 304 (1965)

\bibitem{Buryy2010}
Buryy, O. et al. Thermal changes of the crystal structure and the influence of thermo-chemical annealing on the optical properties of YbAlO$_3$ crystals.  {\it  J. Phys.: Condens. Matter.  \bf 22}, 055902 (2010).

\bibitem{Bonville1978}
Bonville, P., Hodges, J. A., Imbert, P., $\&$ Hartmann-Boutron, F. Spin-spin and spin-lattice relaxation of ${\mathrm{Yb}}^{3+}$ in YbAl${\mathrm{O}}_{3}$, TmAl${\mathrm{O}}_{3}$: Yb, and YAl${\mathrm{O}}_{3}$: Yb and magnetic ordering in YbAl${\mathrm{O}}_{3}$ measured by the M\"ossbauer effect, {\it Phys. Rev. B \bf 18}, 2196 (1978).

\bibitem{Bonville1980}
Bonville, P., Hodges, J. A., $\&$ Imbert, P. Yb$^{3+}$ in RAlO${_3}$ (R = Eu, Gd, Tb, Dy, Ho, Er). A $^{170}$Yb M\"ossbauer effect study of the hyperfine parameters, magnetic ordering and relaxation, {\it J. Physique \bf 41}, 1213 (1980).

\bibitem{Lake2010}
Lake, B. et al. Confinement of Fractional Quantum Number Particles in a Condensed Matter System: Example of CaCu$_2$O$_3$, {\it Nat. Phys, \bf 6}, 5055 (2010).

\bibitem{Rau18}
Jeffrey G. Rau and Michel J. P. Gingras, Frustration and anisotropic exchange in ytterbium magnets with edge-shared octahedra, Phys. Rev. B {\bf 98}, 054408 (2018).

\bibitem{Hester18}
Hester, G. et al. A Novel Strongly Spin-Orbit Coupled Quantum Dimer Magnet: Yb$_2$Si$_2$O$_7$.
arXiv:1810.13096.

\bibitem{Essler97}
Delfino, G., Essler, F. H. L., $\&$ Tsvelik, A. M. Quasi-one-dimensional spin-1/2 Heisenberg magnets in their ordered phase: correlation functions, {\it Phys Rev. B \bf 56}, 11001 (1997).

\bibitem{White1992}
White, S. R. Density matrix formulation for quantum renormalization groups. {\it Phys. Rev. Lett.  \bf 69}, 2863 (1992).

\bibitem{White1993}
White, S. R. Density-matrix algorithms for quantum renormalization groups.  {\it Phys. Rev. B.  \bf 48}, 10345 (1993).

\bibitem{Lake13}
Lake, B. et al. Multispinon Continua at Zero and Finite Temperature in a Near-Ideal Heisenberg Chain. {\it Phys. Rev. Lett.  \bf 111}, 137205 (2013).

\bibitem{Kohgi1997}
Kohgi, M., Iwasa, K., Mignot, J.-M., Ochiai, A. $\&$ Suzuki, T. One-dimensional antiferromagnetic coupling in the low-carrier heavy-electron system Yb$_4$As$_3$: The role of charge ordering, {\it Phys. Rev. B \bf 56}, R11388 (1997)

\bibitem{Kohgi2001}
Kohgi, M. et al. Staggered Field Effect on the One-Dimensional $\mathit{S}\phantom{\rule{0ex}{0ex}}=\phantom{\rule{0ex}{0ex}}\frac{1}{2}$ Antiferromagnet ${\mathrm{Yb}}_{4}{\mathrm{As}}_{3}$, {\it Phys. Rev. Lett. \bf 86}, 2439 (2001)

\bibitem{Caux2005}
Caux, J. -S., Hagemans, R. $\&$ Maillet, J. M. Computation of dynamical correlation functions of Heisenberg chains: the gapless anisotropic regime, {\it J. Stat. Mech.} (2005) P09003.

\bibitem{Caux2012}
Caux, J. -S., Konno, H., Sorrell, M. $\&$ Weston, R. Exact form-factor results for the longitudinal structure factor of the massless XXZ model in zero field, {\it J. Stat. Mech.} (2012) P01007.

\bibitem{Flouquet2005}
Flouquet J. On the Heavy Fermion Road.  {\it  Progress in Low Temperature Physics  \bf 15}, 139 (2005).

\bibitem{Muller1981}
M$\rm \ddot{u}$ller, G., Thomas, H., Beck, H. $\&$ Bonner, J. C. Quantum spin dynamics of the antiferromagnetic linear chain in zero and nonzero magnetic field.  {\it Phys. Rev. B. \bf 24}, 1429 (1981).

\bibitem{Elliott1961}
Elliott, R. J. Phenomenological Discussion of Magnetic Ordering in the Heavy Rare-Earth Metals. {\it Phys. Rev. \bf 124}, 346 (1961).

\bibitem{Fisher1980}
Fisher, M. E. $\&$ Selke, W. Infinitely Many Commensurate Phases in a Simple Ising Model. {\it Phys. Rev. Lett. \bf 44}, 1502 (1980).

\bibitem{Selke1988}
Selke, W. The ANNNI model - Theoretical analysis and experimental application.  {\it Phys. Rep.  \bf 170}, 213 (1988).

\bibitem{Sachdev1999}
Sachdev, S. Quantum Phase Transitions (Cambridge University Press, Cambridge, 1999).

\bibitem{Zapf2014}
Zapf, V., Jaime, M. $\&$ Batista, C. D. Bose-Einstein condensation in quantum magnets. {\it Rev. Mod. Phys. \bf 86}, 563 (2014).

\bibitem{Schroder2000}
Schr{\"o}der, A.  et al. Onset of antiferromagnetism in heavy-fermion metals.  {\it  Nature  \bf 407}, 351 (2000).

\bibitem{Matsumoto2011}
Matsumoto, Y.  et al. Quantum Criticality Without Tuning in the Mixed Valence Compound $\beta$-YbAlB$_4$.  {\it  Science  \bf 331}, 316-319 (2011).

\bibitem{Bianchi2003}
Bianchi, A.  et al. Avoided Antiferromagnetic Order and Quantum Critical Point in
  $\mathrm{C}\mathrm{e}\mathrm{C}\mathrm{o}\mathrm{I}{\mathrm{n}}_{\mathrm{5}}$.  {\it  Phys. Rev. Lett.  \bf 91}, 257001 (2003).

\bibitem{Wu2014}
Wu, L. S. et al. Quantum critical fluctuations in layered YFe$_2$Al$_{10}$.  {\it  Proc. Natl. Acad. Sci. U.S.A.  \bf 111}, 14088-14093 (2014)

\bibitem{Blanc2018}
Blanc, N. et al. Quantum criticality among entangled spin chains. {\it Nat. Phys. \bf 14}, 273 (2018).

\bibitem{Hirobe2016}
Hirobe, D. et al. One-dimensional spinon spin currents.  {\it  Nat. Phys.  \bf 13}, 30 (2016).



\bibitem{Faraday}
Sakakibara T. et al. Faraday Force Magnetometer for High-Sensitivity Magnetization Measurements at Very Low Temperatures and High Fields.  {\it  Jpn. J. Appl. Phys. \bf 33}, 5067 (1994).

\bibitem{Wilhelm2004}
Wilhelm H. et al. A compensated heat-pulse calorimeter for low temperatures.  {\it  Rev. Sci. Instrum. \bf 75}, 2700-2705 (2004).

\bibitem{Ehlers2011}
Ehlers G. et al. The new cold neutron chopper spectrometer at the Spallation Neutron Source: design and performance.  {\it  Rev. Sci. Instrum. \bf 82}, 085108 (2011).

\bibitem{Ehlers2016}
Ehlers G., Podlesnyak A. $\&$ Kolesnikov A. I. The cold neutron chopper spectrometer at the Spallation Neutron Source - A review of the first 8 years of operation.  {\it  Rev. Sci. Instrum. \bf 87}, 093902 (2016).

\bibitem{Dave}
Azuah R. et al. DAVE: a comprehensive software suite for the reduction, visualization, and analysis of low energy neutron spectroscopic data. {\it J. Res. Natl. Inst. Stan. Technol. \bf 114}, 341 (2009).

\bibitem{Mantid}
Arnold O. et al. Mantid -- Data analysis and visualization package for neutron scattering and $\mu$SR experiments. {\it Nucl. Instrum. Methods Phys. Res. Sect. A \bf 764}, 156 (2014).

\end{thebibliography}
\end{document}


\title{Supplementary Information\\ Tomonaga-Luttinger Liquid Behavior and Spinon Confinement in YbAlO$_3$ }

\author{L.~S.~Wu}
\affiliation{Neutron Scattering Division, Oak Ridge National Laboratory, Oak Ridge, TN 37831, USA}
\affiliation{Department of Physics, Southern University of Science and Technology, Shenzhen 518055, China}

\author{S.~E.~Nikitin }
\affiliation{Max Planck Institute for Chemical Physics of Solids, N\"{o}thnitzer Str. 40, D-01187 Dresden, Germany}
\affiliation{Institut f{\"u}r Festk{\"o}rper- und Materialphysik, Technische Universit{\"a}t Dresden, D-01069 Dresden, Germany}

\author{Z.~Wang}
\affiliation{Department of Physics and Astronomy, The University of Tennessee, Knoxville, TN 37996, USA}

\author{W.~Zhu}
\affiliation{Westlake Institute of Advanced Study, Hangzhou, 310024, P.~R.~China}
\affiliation{Theoretical Division, T-4 and CNLS, Los Alamos National Laboratory, Los Alamos, NM 87545, USA}

\author{C.~D.~Batista}
\affiliation{Department of Physics and Astronomy, The University of Tennessee, Knoxville, TN 37996, USA}
\affiliation{Shull-Wollan Center, Oak Ridge National Laboratory, Oak Ridge, TN 37831, USA}

\author{A.~M.~Tsvelik}
\affiliation{Condensed Matter Physics and Materials Science Division, Brookhaven National Laboratory, Upton, NY 11973, USA}

\author{A.~M.~Samarakoon}
\affiliation{Neutron Scattering Division, Oak Ridge National Laboratory, Oak Ridge, TN 37831, USA}

\author{D.~A.~Tennant}
\affiliation{Materials Science and Technology Division, Oak Ridge National Laboratory, Oak Ridge, TN 37831, U.S.A.}
\affiliation{Shull-Wollan Center, Oak Ridge National Laboratory, Oak Ridge, TN 37831, USA}

\author{M.~Brando}
\affiliation{Max Planck Institute for Chemical Physics of Solids, N\"{o}thnitzer Str. 40, D-01187 Dresden, Germany}

\author{L.~Vasylechko}
\affiliation{Lviv Polytechnic National University, 79013 Lviv, Ukraine}

\author{M.~Frontzek}
\affiliation{Neutron Scattering Division, Oak Ridge National Laboratory, Oak Ridge, TN 37831, USA}

\author{A.~T.~Savici}
\affiliation{Neutron Scattering Division, Oak Ridge National Laboratory, Oak Ridge, TN 37831, USA}

\author{G.~Sala}
\affiliation{Neutron Scattering Division, Oak Ridge National Laboratory, Oak Ridge, TN 37831, USA}

\author{G.~Ehlers}
\affiliation{Neutron Technologies Division, Oak Ridge National Laboratory, Oak Ridge, TN 37831, USA}

\author{A.~D.~Christianson}
\affiliation{Materials Science and Technology Division, Oak Ridge National Laboratory, Oak Ridge, TN 37831, U.S.A.}
\affiliation{Neutron Scattering Division, Oak Ridge National Laboratory, Oak Ridge, TN 37831, USA}

\author{M.~D.~Lumsden}
\affiliation{Neutron Scattering Division, Oak Ridge National Laboratory, Oak Ridge, TN 37831, USA}

\author{A.~Podlesnyak}
\affiliation{Neutron Scattering Division, Oak Ridge National Laboratory, Oak Ridge, TN 37831, USA}

\maketitle

{\bf Supplementary Note 1: Specific Heat and Phonon Correction}

The temperature dependent specific heat $C_{\rm{p}}$ of YbAlO$_3$ is shown in Supplementary Fig.~\ref{Cphonon}. A sharp peak anomaly is observed at $T_{\rm{N}}=0.88$ K in zero field, indicating a phase transition into the long-range antiferromagnetic (AFM) order at lower temperatures. By applying a magnetic field along the $a$-axis, the peak is gradually suppressed, and finally disappears in high fields. The lattice contribution of the phonons is estimated through the zero field specific heat of the nonmagnetic isostructural compound YAlO$_3$, shown as empty circles in Supplementary Fig.~\ref{Cphonon}a.
The magnetic specific heat in different fields is then extracted as $C_{\rm{M}}=C_{\rm{p}}-C_{\rm{phonon}}$, as shown in Supplementary Fig.~\ref{Cphonon}b, and all the magnetic specific heat data shown below and in the main text are extracted in the same way.

\begin{figure}[h!]
  \centering
  \includegraphics[width=0.8\columnwidth]{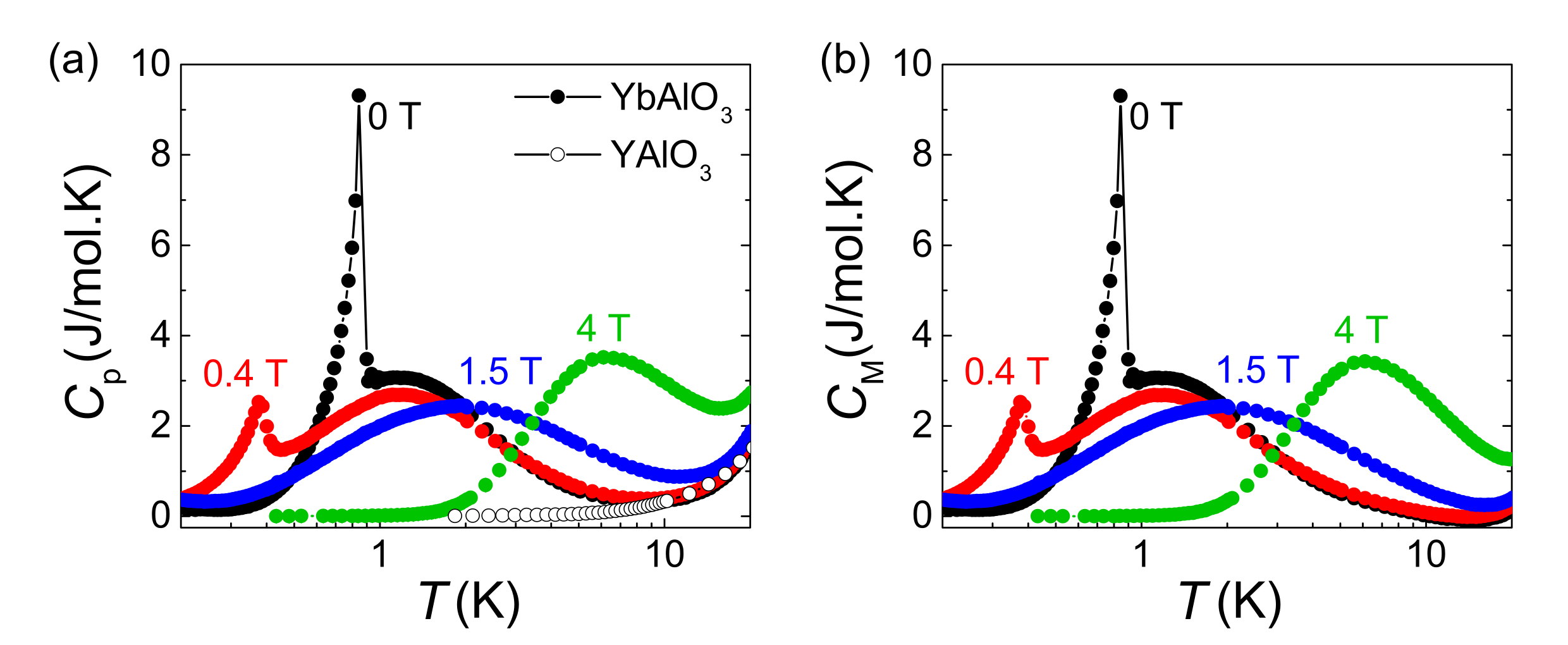}
  \caption{Specific heat and phonon correction. (a) Temperature dependence of specific heat $C_{\rm{p}}$ of YbAlO$_3$ (solid circles) and YAlO$_3$ (empty circles), measured in different magnetic fields applied along the $a$-axis.
  (b) Magnetic specific heat $C_{\rm{M}}$ in different magnetic fields after the correction for the phonon contribution.}
  \label{Cphonon}
\end{figure}

{\bf Supplementary Note 2: Two Magnetic Sublattices and Staggered Field}

\begin{figure}
\centering
\includegraphics[width=0.65\columnwidth]{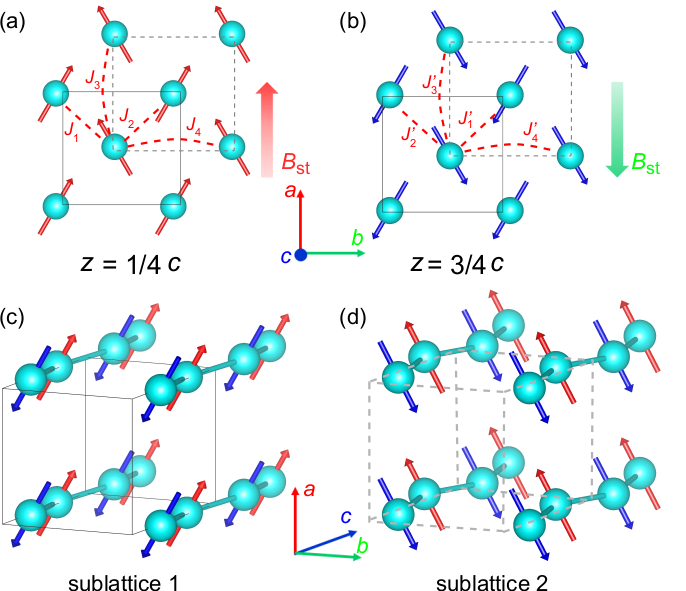}
  \caption{Magnetic structure of YbAlO$_3$.
  (a)-(b) Magnetic Yb moment configurations in planes at $z=1/4c$ (a) and $z=3/4c$ (b), respectively.
  The black solid and gray dashed lines indicate the unit cell for two magnetic sublattices.
  The red dashed lines indicate the near neighbor dipole-dipole interaction between the Yb Ising like moments, and this interaction leads to a net staggered field $B_{\rm{st}}$, with opposite directions in $z=1/4c$ and $z=3/4c$ planes, as indicated by the big red (a) and green (b) arrows.
  (c)-(d) Two separated magnetic sublattices with one-dimensional Yb AFM chains along the $c$-axis, where the Yb moments have the same local Ising axis in each one magnetic sublattice.}
  \label{Mstructure}
\end{figure}

Supplementary Figure~\ref{Mstructure} illustrates the magnetic structure of YbAlO$_3$ in zero field. The Yb moments are confined in the $ab$-plane at $z=1/4c$ (Supplementary Fig.~\ref{Mstructure}a) and $z=3/4c$ (Supplementary Fig.~\ref{Mstructure}b). A static magnetic order is established in YbAlO$_3$ below $T_{\rm{N}}=0.88$~K, and the magnetic configuration of $AxGy$ is selected by the dipole-dipole interaction. As indicated by the red dashed lines, the calculated  dipole-dipole interactions between near neighbors are $J_1=-0.058$~K, $J_2=-0.037$~K, $J_3=-0.10$~K and $J_4=0.03$~K in the $z=1/4c$ plane, assuming a static moment on each Yb site
of 3.8~$\mu_{\rm{B}}{\rm{/Yb}}$. The magnetic moments in $z=3/4c$ are antiparallel to moments in $z=1/4c$, and the magnetic interactions are of the same magnitude, but with opposite signs. In the ordered state, the molecular field produced by these interactions results into a staggered  field $B_{\rm{st}}=-2\sum_{i=1}^4J_{i}\simeq0.33$~K. Since the dipole iteration is long range, we have extended this calculation to about forty-eight near neighbors, and the staggered molecular field saturates to about 0.49 K, which is quite close to the best fit $B_\text{st}=0.27J\sim 0.66$ K used in the DMRG calculation in the main text. With the picture of two sublattices in a staggered field (Supplementary Fig.~\ref{Mstructure}c,d), one can clearly see one dimensioanl (1D)  Yb antiferromagnetic (AFM) chains along the $c$-axis emerging from the three dimensional perovskite crystal structure. This observation naturally leads to the 1D Hamiltonian (1) proposed in the main text.

{\bf Supplementary Note 3: Estimation of interchain interaction}

\begin{figure}
\centering
\includegraphics[width=0.65\columnwidth]{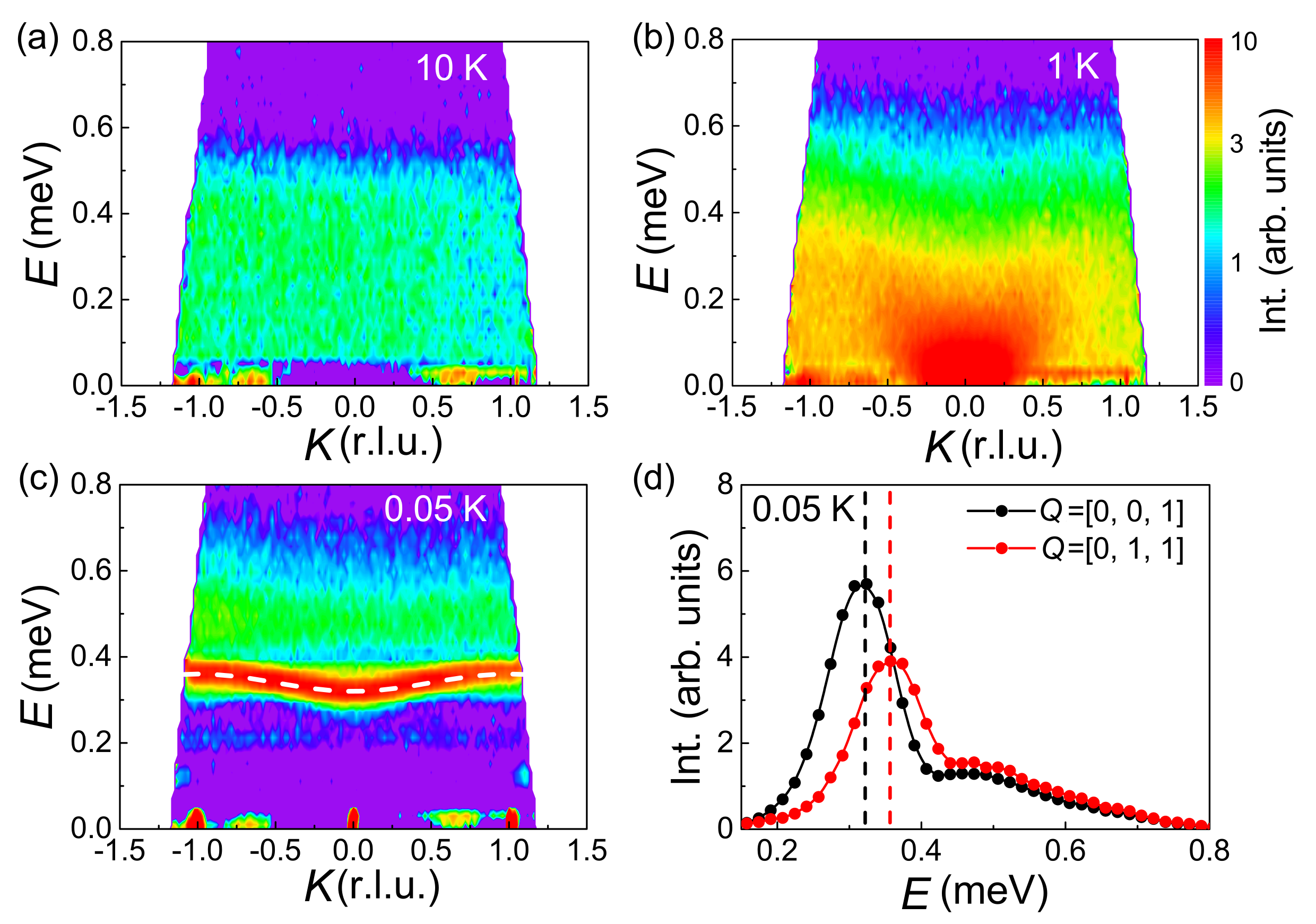}
  \caption{ INS spectra of YbAlO$_3$ along the wave vector (0$K$1) direction.
  (a)-(c) Energy dependent INS spectrum integrated over wave vector $H=[-0.2,0.2]$ r.l.u. and $L=[0.9,1.1]$ r.l.u., measured in zero field and at different temperatures 10~K (a), 1~K (b), and 0.05~K (c), as indicated.
  The dashed line in (c) is described in the text.
  (d) Energy cuts measured at temperature 0.05~K, integrated over wave vector $H=[-0.2,0.2]$ r.l.u., $L=[0.9,1.1]$ r.l.u., $K=[0,0.2]$ r.l.u. and $K=[0.9,1.1]$ r.l.u., respectively.
  The vertical dashed lines indicate the peak positions of 0.32~meV at $\bm{Q}=(0,0,1)$ and 0.36~meV at $\bm{Q}=(0,1,1)$.}
  \label{Kdispersion}
\end{figure}

As expected from the  two-magnetic-sublattice model, where one-dimensional spin chains run along the $c$-axis, we observe a spinon continuum along the  $(00L)$ direction  with only weak modulation in the direction perpendicular to the chain. Supplementary Figure~\ref{Kdispersion}a-c shows the inelastic neutron scattering (INS) spectrum of YbAlO$_3$ at different temperatures along the wave vector direction $(0K1)$.
At $T\sim{10}$~K, well above the magnetic phase transition, thermal fluctuations are large enough to destroy the interchain coupling in the $ab$-plane, and a broad flat continuum is observed (Supplementary Fig.~\ref{Kdispersion}a) with non-observable dispersion along the $(0K1)$ direction.
At $T\sim{1}$~K, i. e. just above the ordering transition, short-range correlations start to build up with broad diffuse scattering developing around $\bm{Q}=(0,0,1)$, and a weak variation is observed in the spectrum (Supplementary Fig.~\ref{Kdispersion}b). Static (ordered) magnetic  moments emerge upon further lowering the temperature down to 0.05~K, which is deep inside the AFM order, and the resulting effective staggered field opens a gap at the AFM wave vector $\bm{Q}=(0,0,1)$ (Supplementary Fig.~\ref{Kdispersion}c).
At this temperature, the lowest peak of the dispersion along $(0K1)$ can be phenomenologically described as
\begin{equation}
  \label{sw}
  E_{\rm{k}}=\sqrt{\left[2J'\left(1-\cos{k\pi}\right)\right]^{2}+E^{2}_{0}},
\end{equation}
where $J'$ is the interchain coupling  and $E_0$ is the gap opened by the staggered field at $L=1$.
Energy cuts taken at $\bm{Q}=(0,0,1)$ and $(0,1,1)$ are shown in Supplementary Fig.~\ref{Kdispersion}d.
With fitting parameters $J'=0.04$~meV, and $E_0=0.32$~meV, the above dispersion relation reproduces the peak positions: 0.32~meV at $(0,0,1)$ and 0.36~meV at $(0,1,1)$. This analysis indicates that the effective interchain interaction is roughly $20\%$ of the intrachain exchange: $J'/J\simeq0.04/0.21\simeq0.19$.

{\bf Supplementary Note 4: Polarization Factor and Longitudinal Fluctuations}

\begin{figure}
  \centering
  \includegraphics[width=0.65\columnwidth]{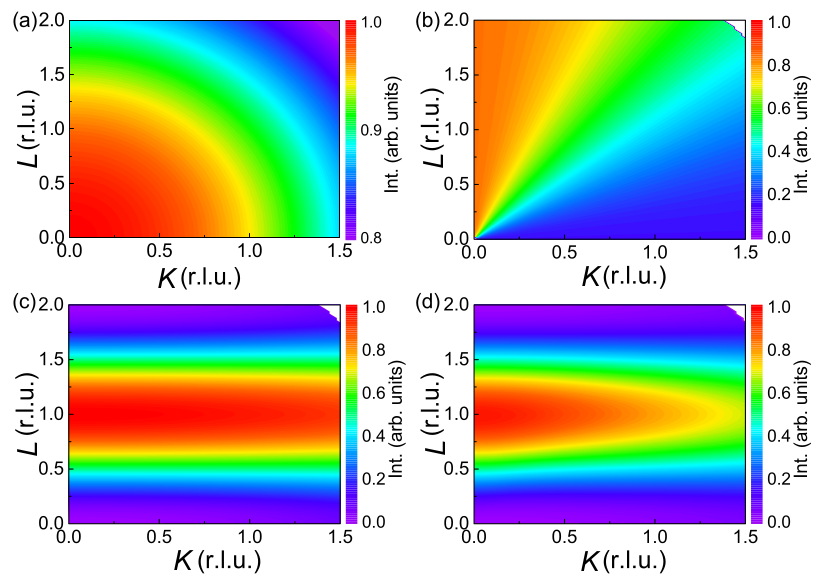}
  \caption{Calculated magnetic neutron scattering factor in the $(0KL)$ scattering plane.
  (a) Contour plot of the magnetic form factor ($|F(\bm{Q})|^{2}$) of Yb$^{3+}$.
  (b) Contour plot of the polarization factor ($1-\hat{Q}_{z}\hat{Q}_{z}$), with the magnetic Ising moments constricted at an angle $\varphi=\pm23.5^{\circ}$ with the $a$-axis.
  (c) Contour plot of the antiferromagnetic static spin structure factor ($S(\bm{Q})$) for nearest neighbor exchange along the $c$-axis.
  (d) Contour plot of the overall cross section (after including the magnetic form factor) (a), polarization factor (b), and the AFM static structure factor (c).}
  \label{Polarization1}
\end{figure}

DC magnetization and CEF studies suggest well separated doublets for the Yb$^{3+}$ ion in YbAlO$_3$.
Since the first exited CEF levels are high (29~meV~$\sim{345}$~K), the low temperature magnetic properties are dominated by the ground state doublets, which can be effectively described as pseudo-spin S=1/2 states with anisotropic $g$-tensors:
\begin{align}\label{geff}
M _{z}=M_{\rm s}= &3.8 \mu_{\rm B}\rm/Yb,&\ \ &g^{ zz}_{\rm eff}=7.6,\\
M _{xy}=M_{\rm c}= &0.23 \mu_{\rm B}\rm/Yb,&\ \ &g^{xx}_{\rm eff}=g^{yy}_{\rm eff}=0.46,
\end{align}
where $z$ is chosen along the local moment easy axis.
This Ising-like $g$-tensor  anisotropy does not necessarily lead to anisotropic interactions,
but it manifests in the magnetic neutron scattering cross-section~\cite{Mourigal2015,Wu2016}:
\begin{equation}\label{Scattering_factor1}
\frac{{\rm d}^{2}\sigma}{{\rm d}\Omega {\rm d}E} \varpropto |F(\bm{Q})|^{2}\sum_{\alpha\beta}(\delta_{\alpha\beta}-\hat{Q}_{\alpha}\hat{Q}_{\beta})(g^{\alpha\beta})^{2}S^{\alpha\beta}(\bm{Q},E).
\end{equation}
Here $|F(\bm{Q})|^{2}$ is the magnetic form factor of Yb$^{3+}$,
$\delta_{\alpha\beta}-\hat{Q}_{\alpha}\hat{Q}_{\beta}$ is the polarization factor,
and $S^{\alpha\beta}(\bm{Q},E)$ is the  dynamical spin structure factor of different components.
Since $(g^{zz})^2/(g^{xx})^2\simeq273$, transverse contributions to the total magnetic scattering can be neglected and the overall magnetic scattering is dominated by  longitudinal fluctuations:
\begin{equation}\label{Scattering_factor2}
\frac{{\rm d}^{2}\sigma}{{\rm d}\Omega {\rm d}E} \varpropto |F(\bm{Q})|^{2}(1-\hat{Q}_{z}\hat{Q}_{z})(g^{zz})^{2}S^{zz}
(\bm{Q},E)+O\left((g^{xx})^2\right)+O\left((g^{yy})^2\right).
\end{equation}
The magnetic polarization factor can be obtained from the angle between the Yb moments and the $a$-axis ($\varphi=\pm23.5^{\circ}$):
\begin{equation}
\label{polarization_factor_equation1}
1-\hat{Q}_{z}\hat{Q}_{z}=1-\frac{Q^2_{k}\rm sin^{2}\varphi}{Q^2_{k}+Q^2_{l}}=
\frac{Q^2_{k}\rm cos^{2}\varphi+Q^2_{l}}{Q^2_{k}+Q^2_{l}}.
\end{equation}
Assuming AFM interactions between nearest-neighbor Yb ions along the $c$-axis, we can also obtain the static magnetic structure factor
\begin{eqnarray}
\label{structure_factor_equation1}
S(\bm{Q})=\int S(\bm{Q},E){\;}{\rm d}E =|\sum_{\rm{i}} \bm{m}_{\rm{i}}e^{-i\bm{Q}\cdot \bm{r}_{\rm{i}}}|^{2} \nonumber\\
=2|m|^{2}\cdot\left[\sin^{2}\frac{\pi(l+0.12k)}{2}+\sin^{2}\frac{\pi(l-0.12k)}{2}\right],
\end{eqnarray}
where $\bm{m}_{\rm{i}}$ are the Yb magnetic moments located at positions $\mathbf{r}_{\rm{i}}$~\cite{Igor2015}.  The Yb atoms form into a zig-zag chain along the crystal $c$-axis in the orthorhombic perovskite structure, with small distortions in $a$ and $b$ directions (Supplementary Fig.~\ref{Mstructure}). The distortion angle along the $a$-axis is about $\sim2.2^{\circ}$, which is  negligible for scattering in the $(0KL)$ plane. The larger distortion angle $\sim9.9^{\circ}$ along the $b$-aixs  introduces an additional wave vector dependence with $l\pm0.12k$ in the scattering factor, making the spectral weight around $L=1$ more spread at higher values of wave vector $K$.

The calculated magnetic form factor ($|F(\bm{Q})|^{2}$), polarization factor ($1-\hat{Q}_{z}\hat{Q}_{z}$) and AF  static spin structure factor ($S(\bm{Q})$) are shown in Supplementary Fig.~\ref{Polarization1}a-c. The overall magnetic scattering factor is plotted in Supplementary Fig.~\ref{Polarization1}d. The calculated wave vector dependence of the magnetic scattering can be directly compared to the experimental results, as presented in Supplementary Fig.~\ref{Polarization2}.

\begin{figure}[!tbp]
	\centering
	\includegraphics[width=0.75\columnwidth]{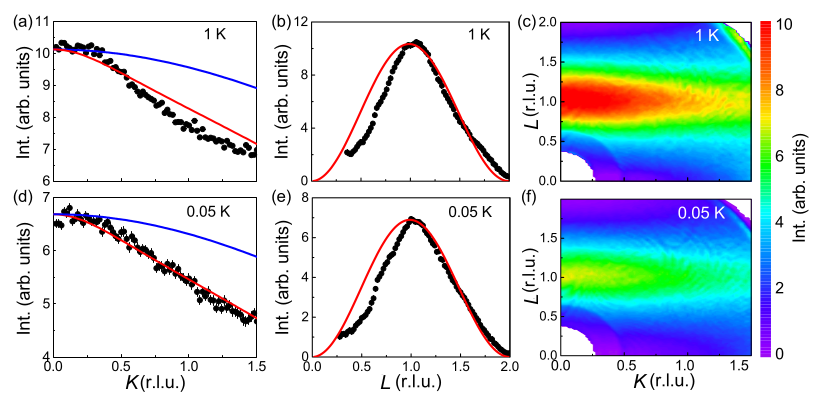}
	\caption{Experimental magnetic INS spectrum of YbAlO$_3$ in the $(0KL)$ plane.
		(a),(d) Constant energy cut integrated over $E=[0.1,0.8]$~meV, and wave vector $H=[-0.2,0.2]$ r.l.u., $L=[0.9,1.1]$ r.l.u., measured at temperatures 1~K (a), and 0.05~K (d), respectively.
		The blue lines are the calculated form factor ($|F(\bm{Q})|^{2}$) only, and the red lines are the product of the calculated form and polarization factors.
		(b),(e) Constant energy cut integrated over energy $E=[0.1,0.8]$~meV, and wave vector $H=[-0.2,0.2]$ r.l.u., $K=[-0.2,0.2]$ r.l.u., measured at temperatures 1~K (d), and 0.05~K (e), respectively.
		The red lines are the calculated magnetic structure factor.
		(c),(f) Contour plots of the INS spectrum in the $(0KL)$ scattering plane, integrated over energy $E=[0.1,0.8]$~meV, at temperatures 1~K (c), and 0.05~K (f), respectively.}
	\label{Polarization2}
\end{figure}

The wave vector dependence along the ($0K1$) direction, measured at temperatures 1~K and 0.05~K, is shown in Supplementary Fig.~\ref{Polarization2}a,d, respectively. The blue lines are the calculated form factor ($|F(\bm{Q})|^{2}$), which changes about 10\% along the $K$ direction, while the additional 20\% changes are well captured when the polarization factor is included (red lines).  The polarization factor is a constant  for $K=0$ and the intensity variation along the $(00L)$ direction is dominated by the spin structure factor $S(\bm{Q})$. As shown in Supplementary Fig.~\ref{Polarization2}b,e, both the  $L$-dependence below and above the AFM phase transition are well described by the calculated magnetic structure factor (red lines). The experimental INS spectrum integrated over $E=[0.1,0.8]$~meV in the $(0KL)$ scattering plane is shown in Supplementary Fig.~\ref{Polarization2}c,f, where an overall consistency is observed with the calculated pattern shown in Supplementary Fig.~\ref{Polarization1}d. This demonstrates that the magnetic scattering is dominated by longitudinal fluctuations along the local easy-axis of the Yb moments.

{\bf Supplementary Note 5: Comparison of paramagnon and two-spinon continuum}

\begin{figure}
	\centering
	\includegraphics[width=0.8\columnwidth]{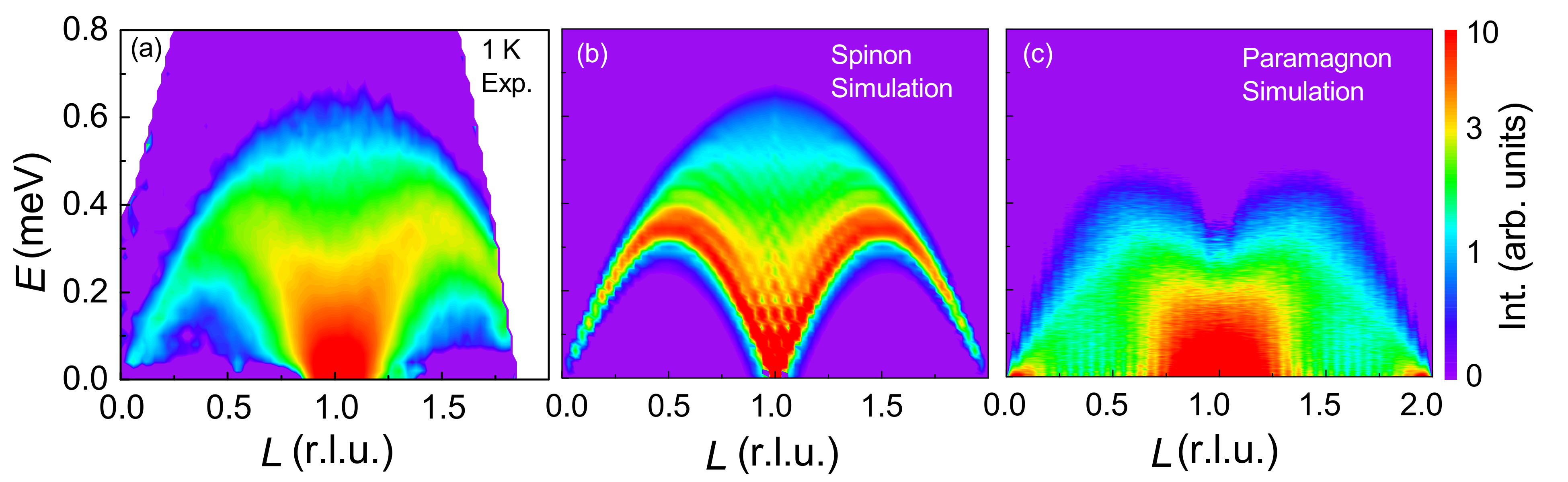}
	\caption{Comparison of the spin excitation spectra.
      (a) Experimental INS spectrum at 1 K (the experimental N\'eel temperature is $T_N\approx 0.88$ K).
      (b)  Longitudinal spin structure factor $S^{zz}(\bm{Q},E)$ of model Eq.~(2) in the main text, computed by DMRG at $T=0$.
      (c) Longitudinal spin structure factor $S^{zz}(\bm{Q},E)$ of the classical spin model~\eqref{eq:classical} at 0.5 K (the corresponding N\'eel temperature is $T_N\approx 0.45$ K), obtained from the Landau-Lifshitz dynamics (the energy $E$ has been rescaled by $\sqrt{3}/2$ to account for the size of $S=1/2$). \label{fig:paramagnon}}
\end{figure}

\begin{figure}
	\centering
	\includegraphics[width=0.8\columnwidth]{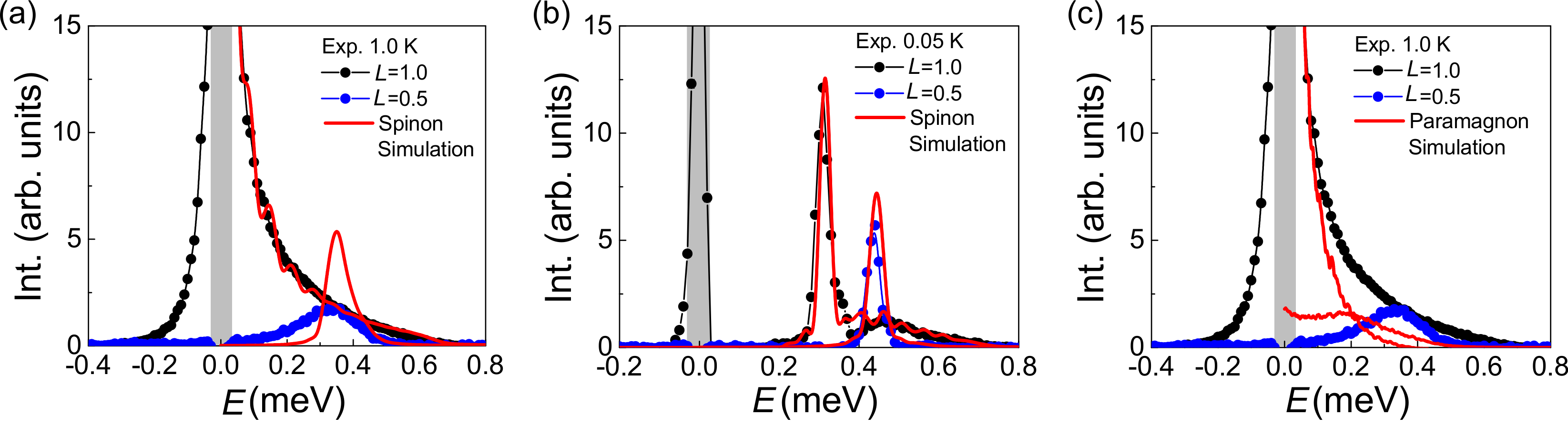}
	\caption{ Energy cuts of the spin excitation spectrum.
             (a) Energy cut at wave vector $H=[-0.2, 0.2]$ r.l.u., $K=[-1, 1]$ r.l.u., $L=[0.9, 1.1]$ r.l.u., and $L=[0.4, 0.6]$ r.l.u. at 1.0 K, and the red line is the
             energy cut of $S^{zz}(\bm{Q},E)$ from $T=0$ DMRG calculation with zero staggered field $B_{\rm st}=0$.
             (b) Energy cut at wave vector $H=[-0.2, 0.2]$ r.l.u., $K=[-1, 1]$ r.l.u., $L=[0.9, 1.1]$ r.l.u., and $L=[0.4, 0.6]$ r.l.u. at 0.05 K, and the red line is the
             energy cut of the $T=0$ DMRG calculation with staggered field $B_{\rm st}/J=0.27$.
             In both (a) and (b), the simulated intensity are scaled by the same factor to match the experimental data
             (the artificial oscillations in the simulated curve are from the finite size effect).
             (c) Energy cut at wave vector $H=[-0.2, 0.2]$ r.l.u., $K=[-1, 1]$ r.l.u., $L=[0.9, 1.1]$ r.l.u., and $L=[0.4, 0.6]$ r.l.u. at 1.0 K, and the red line is the
              energy cut of $S^{zz}(\bm{Q},E)$ from Landau-Lifshitz spin dynamics at $T=0.5$~K.
             The vertical gray bar indicates the instrumental resolution, which is much narrower than the experimental spinon continuum.
              \label{fig:paramagnoncut}}
\end{figure}

In the main text, we have shown that the magnetic excitations of YbAlO$_3$ have a two-spinon continuum at $T=1$~K (Fig.~3a), right above the N\'eel temperature $T_N$. The spectrum is consistent with our $T=0$ DMRG calculation (Fig.~3c). Furthermore, the broadening of the experimental data can also be captured by a finite temperature tDMRG calculation~\cite{Lake2013}.

While everything is consistent with the two-spinon continuum, it is still interesting to ask if one can obtain the same excitation spectrum from the paramagnon picture. For this comparison, we use the 3-dimensional version of the effective model, with classical spins:
\begin{equation}\label{eq:classical}
	\mathcal{H} = J \sum_{\langle ij \rangle} \bm{S}_i \cdot \bm{S}_j + J_\text{inter} \sum_{\langle ij \rangle^\prime} S_i^z S_j^z,
\end{equation}
where $\langle ij \rangle$ denotes the nearest neighbor bonds along the chain, and $\langle ij \rangle^\prime$ denotes the nearest neighbor bonds in between chains. Following the best fit from the main text, we use $J=0.21$~meV, $J_\text{inter}\approx \frac{B_\text{st}}{2} \approx 0.028$~meV.

To calculate the excitation spectrum of the classical spin model \eqref{eq:classical}, we first use Monte-Carlo simulation with Metropolis update to obtain the spin configurations in thermal equilibrium, on a $6\sqrt{2}\times 6\sqrt{2}\times 40$ lattice (40 being along the chain direction). After achieving thermalization from completely disordered spin configurations, we time-evolve the spins with Landau-Lifshitz dynamics, using in total $3\times10^4$ dynamical steps. The longitudinal component of structure factor is obtained by Fourier transformation of the spin correlation function $\langle S^z(\bm{r}_i,0) S^z(\bm{r}_j,t) \rangle$, averaged over 32 independent replicas.

Supplementary Figures~\ref{fig:paramagnon} and \ref{fig:paramagnoncut} show the comparison between the experimental INS spectrum and theory. While all of them show broad continuum centered around $\bm{Q}=(0,0,1)$, the details are clearly different: away from $\bm{Q}=(0,0,1)$, the DMRG calculation captures correctly the coherent excitations (Supplementary Fig.~\ref{fig:paramagnoncut}(a),(b)), while the Landau-Lifshitz simulation produces diffuse-like scattering at low energy (Supplementary Fig.~\ref{fig:paramagnoncut}(c)), i.e. a simple paramagnon picture does not apply when the temperature is right above $T_\text{N}$.

{\bf Supplementary Note 6: Quantum critical scaling}

\begin{figure}[!tbp]
	\centering
	\includegraphics[width=0.75\columnwidth]{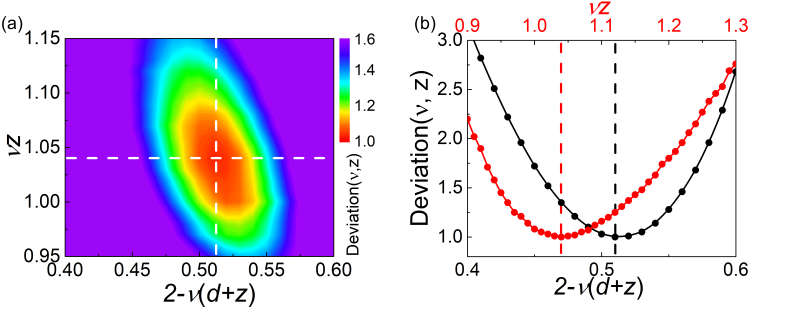}
	\caption{Quantum critical scaling of YbAlO$_3$. (a) Contour plot of the calculated deviation from scaling as functions of $\nu{z}$ and $2-\nu(d+z)$, using the magnetization data (see Fig.~7a-b in the main text). The white dashed lines indicate the global minimum at $\nu{z}=1.04$ and $2-\nu(d+z)=0.51$. (b) Calculated deviations for different scaling parameters around the minimum position.}
	\label{Scalingplot}
\end{figure}

The field dependence of the magnetization is presented in Fig.~7a in the main text.
Following Ref.~\onlinecite{Yang2017}, we define the deviation from scaling as:
\begin{equation}\label{err}
{\rm Deviation}(\nu, z)=\sum_{i,j}\sum_{i',j'}[\frac{{\rm d}M}{{\rm d}B}(T_{\rm i}, B_{\rm j})-\frac{{\rm d}M}{{\rm d}B}(T_{\rm i'}, B_{\rm j'})]^{2}.
\end{equation}

As shown in Supplementary Fig.~\ref{Scalingplot}a, the minimum deviation is reached for
\begin{gather}\label{vz}
 \nu z=1.04\\
 2-\nu(d+z)=0.51,
\end{gather}
which is consistent with a  free fermion fixed point: $\nu=1/2$, $z=2$ and $d=1$.